\documentclass[twocolumn]{aastex701}

\newcommand{\clagnrate}{$9.6^{+4.9}_{-3.4}\%$ }
\newcommand{\syonerate}{$1.2^{+0.87}_{-0.50}\%$ }
\newcommand{\sytworate}{$\leq 0.39\%$ }
\newcommand{\drwrate}{$1.6^{+0.19}_{-0.17}\%$ }

\newcommand{\completeness}{$88^{+5.1}_{-8.0}\%$}

\usepackage{siunitx}
\DeclareSIUnit\angstrom{\text {Å}}
\usepackage{amsmath}

\begin{document}

\title{Identifying Changing-Look AGN Transitions in Light Curve Data with the Zwicky Transient Facility}

\author[0000-0003-1535-4277]{Margaret E. Verrico}
\affiliation{University of Illinois Urbana-Champaign Department of Astronomy, University of Illinois, 1002 W. Green St., Urbana, IL 61801, USA}
\affiliation{Center for AstroPhysical Surveys, National Center for Supercomputing Applications, 1205 West Clark Street, Urbana, IL 61801, USA}
\email{verrico2@illinois.edu}
\author[0000-0002-4235-7337]{K. Decker French}
\affiliation{University of Illinois Urbana-Champaign Department of Astronomy, University of Illinois, 1002 W. Green St., Urbana, IL 61801, USA}
\email{deckerkf@illinois.edu}
\author[0000-0003-4703-7276]{Vivienne F. Baldassare}
\affiliation{Department of Physics and Astronomy, Washington State University, Pullman, WA 99163, USA;}
\email{vivienne.baldassare@wsu.edu}
\author[0000-0001-9947-6911]{Colin J. Burke}
\affiliation{Department of Astronomy, Yale University, New Haven, CT 06511, USA}
\email{colin.j.burke@yale.edu}
\author[0000-0003-1752-679X]{Laura Duffy}
\affiliation{Department of Astronomy and Astrophysics and Institute for Gravitation and the Cosmos, The Pennsylvania State University, 525 Davey Lab, University Park, PA
16803, USA}
\email{lrd48@psu.edu}
\author[0000-0003-1714-7415]{Nicholas Earl}
\affiliation{University of Illinois Urbana-Champaign Department of Astronomy, University of Illinois, 1002 W. Green St., Urbana, IL 61801, USA}
\email{nmearl2@illinois.edu}
\author[0009-0003-8161-5057]{Megan Harrison}
\affiliation{University of Illinois Urbana-Champaign Department of Astronomy, University of Illinois, 1002 W. Green St., Urbana, IL 61801, USA}
\email{mmh@illinois.edu}
\author[0000-0001-9668-2920]{Jason T. Hinkle}
\altaffiliation{NHFP Einstein Fellow}
\affiliation{University of Illinois Urbana-Champaign Department of Astronomy, University of Illinois, 1002 W. Green St., Urbana, IL 61801, USA}
\affiliation{NSF-Simons AI Institute for the Sky (SkAI), 172 E. Chestnut St., Chicago, IL 60611, USA}
\email{jhinkle6@illinois.edu }
\author[0000-0002-9158-5408]{Alexander Messick}
\affiliation{Department of Physics, Villanova University, 800 E. Lancaster Ave, Villanova, PA 19085, USA}
\email{amessick@villanova.edu}
\author[0000-0001-9042-965X]{Samaresh Mondal}
\affiliation{University of Illinois Urbana-Champaign Department of Astronomy, University of Illinois, 1002 W. Green St., Urbana, IL 61801, USA}
\email{samaresh@illinois.edu}
\author[0009-0004-0436-0932]{Yashasvi Moon}
\affiliation{University of Illinois Urbana-Champaign Department of Astronomy, University of Illinois, 1002 W. Green St., Urbana, IL 61801, USA}
\email{ymoon@illinois.edu}
\author[0009-0005-1158-1896]{Margaret Shepherd}
\affiliation{University of Illinois Urbana-Champaign Department of Astronomy, University of Illinois, 1002 W. Green St., Urbana, IL 61801, USA}
\email{ms169@illinois.edu}
\author[0000-0002-8501-3518]{Zachary Stone}
\affiliation{University of Illinois Urbana-Champaign Department of Astronomy, University of Illinois, 1002 W. Green St., Urbana, IL 61801, USA}
\email{stone28@illinois.edu}

\date{\today}

\submitjournal{ApJ}

\correspondingauthor{Margaret E.  Verrico}
\email{verrico2@illinois.edu}

\begin{abstract}

Changing-Look AGN (CL-AGN) are AGN which transition between Seyfert types, challenging AGN unification models. Most CL-AGN have been identified via repeat spectroscopy, making it difficult to determine the duration and magnitude of the CL-AGN transition. As such, the physical mechanisms behind this transition are still unknown. We use synthetic photometry in combination with ZTF light curve data to develop a new criterion to identify photometric CL-AGN transitions based on changes in $g$-band magnitude and $g-r$ color. We find that a CL-AGN criterion of $|\Delta g| > 0.4$ mag and $|\Delta(g-r)| > 0.2$ mag recovers a photometric transition in \clagnrate of CL-AGN hosts over the six-year ZTF survey, including a candidate repeating changing-look event in SDSS J084957.78+274728.9. Using simulated AGN light curves, we estimate the false positive rate among the simulated Seyferts to be \drwrate. We find that the rate of similar flares  among Type 1 Seyferts is \syonerate, and among Type 2 Seyferts is \sytworate over six years. Photometric CL-AGN transitions last between 21 and 560 days, with a median duration of 360 days, consistent with the thermal or orbital timescales for AGN disks. We do not detect a correlation between black hole mass and transition duration, likely due to the small sample of detected photometric transitions. This method can be applied to the upcoming Legacy Survey of Space and Time to identify CL-AGN candidates and test theories of their origins.

\end{abstract}


\section{Introduction} \label{sec:intro}

Active galactic nuclei (AGN) are rapidly-accreting supermassive black holes at the centers of galaxies. AGN with broad Balmer emission lines are called ``Type 1" AGN, while AGN which lack broad line emission are called ``Type 2" AGN \citep[][]{Khachikian1974}. These observed differences are explained under the AGN unification model as a difference in orientation; in Type 2 AGN, the orientation of the viewer relative to the dusty torus obscures the virialized clouds close to the black hole which form the Broad Line Region (BLR) \citep[][]{Antonucci1993,urry1995}.  Meanwhile, face-on Type 1 AGN allow for observations of the central accretion disk, which exhibits variable emission at all wavelengths \citep[for a recent review, see][]{paolillo2025}.

AGN variability in the UV/optical may come from accretion rate changes \citep[e.g.][]{Lyubarskii1997}, changes in accretion disk properties \citep[e.g.][]{kelly2009,burke2021}, reprocessing of high-energy photons \citep[e.g.][]{Kazanas_2001, cackett2007,Kammoun2021}, microlensing \citep[e.g.][]{Vernardos2024}, or some combination of all of these \citep{paolillo2025}.  Generally, variability amplitude is anticorrelated with the AGN's luminosity \citep[e.g.][]{wilhite2008,zuo2012,simm2016,Chanchaiworawit2024} and can be modeled as a Continuous Auto-Regressive Moving Average (CARMA) process, either of first order \citep[damped random walk; e.g.][]{kelly2009,burke2021,stone2022}, or second order \citep[damped harmonic oscillator; e.g.][]{yu2022}. None of these models capture all observed AGN variability properties \citep[see e.g.][]{moreno2019}, and in recent years, AGN with variability outside the normal distribution have been dubbed ``extremely variable quasars" \citep[EVQs; e.g.][]{rumbaugh2018}, the most extreme subset of which are changing-look AGN (CL-AGN; \citealt{matt2003,lamassa2015}, and see recent reviews by \citealt{ricci2022} and \citealt{komossa2024}).

CL-AGN are a subset of AGN which have been observed to transition between AGN types on human timescales, challenging traditional viewing angle-dependent AGN unification models. CL-AGN can be classified as changing-obscuration (in which an obscuring medium passes in front of the X-ray corona and the change in AGN type is a change in Compton thickness) or changing-state (in which an intrinsic change in the accretion rate changes the appearance of the broad lines, leading to a transition between Type 1 and 2). Optical CL-AGN usually are not associated with a change in Compton thickness or reddening, implying that a change in obscuration is not sufficient to explain their changing spectral features \citep[see section 3.4 of][and references therein]{ricci2022}. Instead, optical CL-AGN are typically considered to be changing-state AGN and are the focus of this study. The physical cause for this state change is not currently known, but theories include tidal disruption events in AGN disks \citep[e.g.][]{wangyihan2024a}, changes in the fueling state to the AGN \citep[e.g.][]{liu2021,wangj2024}, and instabilities in the accretion flow \citep[e.g.][]{noda2018,sniegowska2020}. High-cadence spectroscopy is generally needed to probe the physics driving anomalous astrophysical transients \citep[e.g.][]{hinkle2022,earl2025}. However, this type of study has been limited for CL-AGN, as it is rare to catch CL-AGN in the early stages of their $\sim$years-long state changes. Very few CL-AGN have ever been monitored through a state change with high-cadence spectroscopy \citep[e.g.][]{wang2018,Trakhtenbrot2019,zeltyn2022, neustadt2023,duffy2025}, and as such we have very little insight into the physical changes occurring in most CL-AGN. Fortunately, the emergence of photometric surveys and survey instruments like the Asteroid Terrestrial-impact Last Alert System \citep[ATLAS;][]{atlas}, the All Sky Automated Survey for SuperNovae \citep[ASAS-SN;][]{asassn}, the Catalina Real-time Transient Survey \citep[CRTS;][]{Djorgovski2011}, the Panoramic Survey Telescope \& Rapid Response System \citep[Pan-STARRS;][]{chambers2016}, the Young Supernova Experiment \citep[YSE;][]{jones2021}, and the Zwicky Transient Facility \citep[ZTF;][]{ztf, ztf2, ztf3}, as well as the upcoming Legacy Survey of Space and Time \citep[LSST;][]{ivezic2019} at the Vera C. Rubin Observatory, offer an opportunity to identify and study large samples of CL-AGN via photometry.

Several methods have already been tested to identify CL-AGN using photometry. \citet{lopeznavas2022}, \citet{wang2024}, and \citet{amrutha2024} identified CL-AGN by searching for AGN with variability characteristics which are not consistent with their archival spectral classification. \citet{hon2022} searched for objects which have galaxy-like or QSO-like colors inconsistent with their archival classification. \citet{yang2018}, \citet{shen2021} \citet{graham2020} select objects based on extreme variability in the optical and mid-infrared when compared to other Seyferts and/or quasars. \citet{zhu2025} used a characteristic bluer-when-brighter flare pattern to identify turn-on CL-AGN with ZTF. Finally, \citet{Frederick2019}, \citet{zheng2024}, and \citet{duffy2025} all searched for atypical photometric rises in Type 2 AGN or Low Ionization Nuclear Emission-line Regions (LINERs) which may indicate a sudden state change. In this work, we synthesize previous work focusing on brightness and color changes to try to identify the changing-look transition in known CL-AGN. The duration and magnitude of this transition should be set by the physical mechanism driving the transition. Thus, if we can identify the optical photometric signatures of CL-AGN state changes, we can use large upcoming surveys like LSST to constrain the physics behind their behavior. 

In this work, we use light curves from the Zwicky Transient Facility (ZTF) to develop a photometric criterion for CL-AGN transitions in a sample of 52 CL-AGN, as well as 260 Seyfert 1 and 257 Seyfert 2 AGN from the catalog of \citet{veroncetty2010} (described in more detail in Section \ref{sec:data}). We describe our light curve and spectral fitting in Section \ref{sec:methods}. We develop a criterion to identify CL-AGN in Section \ref{sec:developing_criterion}. We describe candidate CL-AGN transitions identified in ZTF in Section \ref{sec:clagn_in_ztf}. Finally, we compare to theories for CL-AGN origins in Section \ref{sec:clagn_origins}.

Throughout this work, we assume a flat $\Lambda$CDM cosmology with $H_0 = 70$ km/s/Mpc and $\Omega_m = 0.3$. We use the ZTF AB magnitude system throughout this work.

\section{Data \& Samples} \label{sec:data}

CL-AGN have been discovered serendipitously \citep[e.g.][]{collinsouffrin1973,storchibergmann1993,eracleous2001}, selected based on a change in spectroscopic classification between surveys or between epochs of one survey \citep[e.g.][]{ruan2016, yang2018}, via changes in measured Balmer line width or flux between observations \citep[e.g.][]{yu2020, tozzi2022, dong2024, zeltyn2024}, and/or with triggered follow-up based on photometric changes \citep[e.g.][]{gezari2017, Frederick2019, Trakhtenbrot2019,hon2022,lopeznavas2022,wang2023,wang2024,zhu2025}. These methods are sometimes combined with or confirmed by visual inspection \citep[e.g.][]{green2022, guo2024a}. Each of these methods introduces a bias in selection (e.g. identifying objects based on a photometric flare preferentially selects turn-on CL-AGN). We choose to combine as many CL-AGN from the literature as possible with light curves in ZTF to reduce the impact of selection effects on our results. 

The most dramatic photometric change associated with CL-AGN transitions is the change in AGN continuum blue-ward of the 4000 \si{\angstrom} break, particularly the continuum blue-ward of 3700 \si{\angstrom}. This feature is redshifted into the ZTF $g-$ band at $z=0.13$, limiting the effectiveness of photometric detection with current surveys at low redshift. Meanwhile, this feature is not redshifted into the ZTF $r$-band until $z\sim0.5$, meaning changes in $r$-band should be less extreme. We restrict our sample to $z\geq 0.2$ to allow for sufficient coverage of the AGN continuum in the ZTF $g$ band. For the sake of comparison to the \citet{veroncetty2010} catalog, we also restrict our sample to objects at $z \leq0.4$. 

We begin by collecting CL-AGN catalogs and single-object studies from the literature which contain CL-AGN with observed transitions in H$\alpha$ (here defined as a transition between Type 1 and Type 1.8--2) in the redshift range $0.2 \leq z \leq 0.4$. These sources include \citet{lamassa2015}, \citet{ruan2016}, \citet{runnoe2016}, \citet{gezari2017}, \citet{yang2018}, \citet{green2022}, \citet{guo2024a}, and \citet{zeltyn2024}, which contain a total of 304 CL-AGN. We select only objects classified based on changing-look transitions in H$\alpha$, limiting us to 276 objects. After limiting to $0.2 \leq z \leq 0.4$, we have 70 CL-AGN.

We perform our analysis on light curves from the Zwicky Transient Facility (ZTF). ZTF is a wide-field camera in the Northern hemisphere with a roughly 3-day cadence and median 5$\sigma$ limiting magnitudes of $g\sim20.8$ and $r\sim20.6$ \citep[][]{ztf, ztf2, ztf3}. Importantly, the ZTF $g, r,$ and $i$ bands cover similar wavelength ranges to the LSST $g, r,$ and $i$ bands \citep{ztf2,ivezic2019}, so we can use ZTF light curves to motivate future CL-AGN science with LSST. We elect to focus on the $g$ and $r$ bands both due to the anticipated change blueward of the Balmer break and because only 29 objects in our sample have sufficient $i-$band data. We query ZTF Data Release 23 $g$- and $r$-band light curves \citep[][]{ztf_light_curves} within $3''$ of the literature coordinates of each CL-AGN. 64 objects have both a $g$- and $r$-band light curve in ZTF. We first remove all light curve points with a signal to noise ratio less than 10. We then use Astropy's in-built \texttt{stats.sigma\_clip} function to sigma-clip each light curve at three standard deviations for one iteration to remove spurious points.  After cleaning the data, we require ZTF light curves to contain at least 100 data points for our analysis, leading to a final sample of 52 CL-AGN (see Table \ref{tab:sampleselection}).

\begin{table}[]
    \centering
    \caption{Sample selection for CL-AGN with ZTF data.}
    \begin{tabular}{l|c}
        \textbf{Criterion} & \textbf{Num. CL-AGN} \\
        \hline
        \hline
        Parent sample & 304 \\
        \hline
        H$\alpha$ transition & 276 \\
        \hline
        $0.2\leq z\leq0.4$ & 70 \\
        \hline
        ZTF $g-$ and $r$-band & 64 \\
        \hline
        $\geq100$ data points after cleaning & 52 \\
        \hline
        \hline
        \textbf{Final} & \textbf{52}
    \end{tabular}
    \label{tab:sampleselection}
\end{table}

In Section \ref{sec:J0849}, we additionally use data from the Catalina Real Time Survey \citep[CRTS;][]{Djorgovski2011} to analyze changes in AGN-like variability over time for a candidate repeating CL-AGN. CRTS is a single-band wide-field survey with a limiting magnitude of $\sim20-21$ mag per exposure. We query CRTS data with a $3''$ aperture and similarly use only data points with signal to noise greater than 10 and sigma-clip  at three standard deviations for one iteration.

We select Seyfert 1 and 2 comparison samples from \citet{veroncetty2010} to determine the rate of photometric CL-AGN transitions in ``normal" AGN. In particular, we are interested in the rate of photometric transitions among Seyfert 1s, as Seyfert 2 transitions should be obscured by the dusty torus except in the case of ``true Seyfert 2s," or unobscured AGN which lack a broad line region \citep[][]{tran2001, elitzur2009, elitzur2014}. We construct two redshift bins [0.2, 0.3, 0.4] and randomly select five comparison Seyfert 1 and Seyfert 2 per CL-AGN from each redshift bin, the maximum number of objects per CL-AGN with sufficient data in ZTF. In one bin (the [0.3, 0.4] bin for Seyfert 2s), we are limited by the number of Seyfert 2s with sufficient data and select three fewer comparison objects for this bin. Even with these three objects removed, we retain a sample of Seyfert 2s which is statistically indistinct in its redshift distribution from the CL-AGN with a p-value of 0.37 under a Kolmogorov-Smirnov two-sample test. This leaves us with 260 Seyfert 1 and 257 Seyfert 2 AGN.

\section{Methods} \label{sec:methods}

\subsection{Light curve fitting} \label{sec:light_curve_fitting}

\begin{figure}
    \centering
    \includegraphics[width=\linewidth]{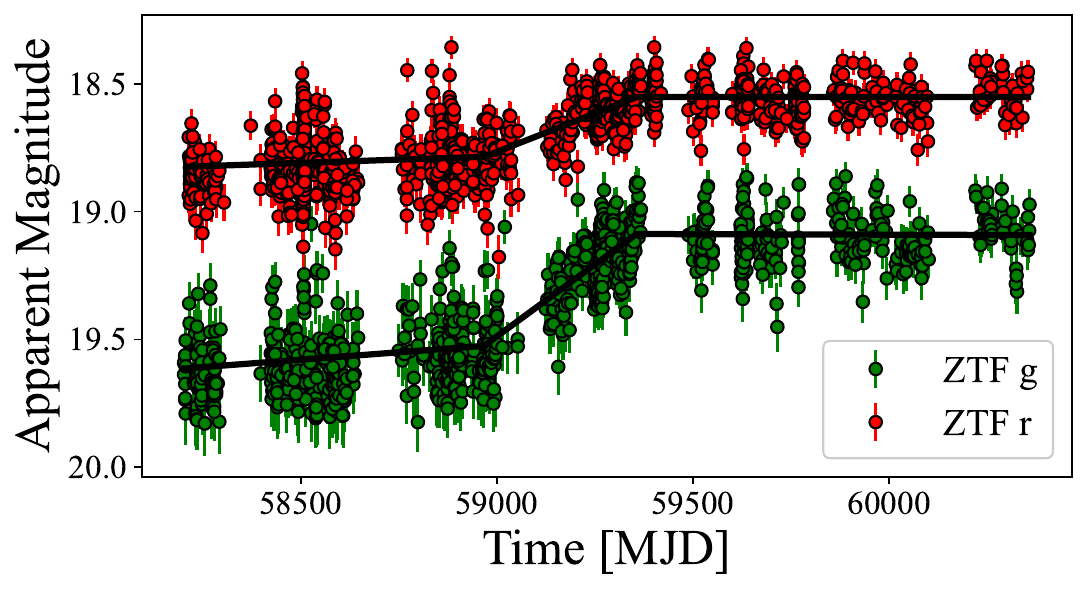}
    \caption{An example three-segment piecewise fit to the light curve of J1011+5442, first identified  as a CL AGN by \citet{runnoe2016}. The above transition was identified as a repeating turn-on CL-AGN by \citet{yang2024,duffy2025,lyu2025}, and \citet{wang2025}. We use \texttt{\texttt{pwlf}} for the fit, where the line breaks in the $r$-band are fixed to the values fit for the $g$-band. Throughout this paper, $\Delta(g)$ refers to the change in magnitude over the center segment of the fit, and $\Delta(g-r)$ measures the same for $g-r$ color.}
    \label{fig:examplefit}
\end{figure}

We then use \texttt{Piecewise Linear Fit} \citep[\texttt{\texttt{pwlf}}; ][]{pwlf} to fit a three-segment piecewise function to each light curve to determine where the CL-AGN transition begins and ends (see example fit in Figure \ref{fig:examplefit}). \texttt{pwlf} uses multiple least-squares fits to the data to determine the optimal break points in an n-segment linear fit. For the purposes of this work, we are interested in three-segment sigmoid-like transitions similar to those found in \citet{duffy2025} to accompany changing-look transitions in three changing-look quasars. These transitions are not easily confused with the light curves of transient flares like TDEs or supernovae.

We compute bootstrapped errors for CL-AGN and CL-AGN candidates (see Section \ref{sec:clagn_candidates}) by sampling a random 90\% of the light curve 100 times per object, then rerunning the fit. Error bars shown on histograms and plots are the 16th and 84th percentiles of the results of this bootstrapping.

\subsection{Black Hole Mass Measurements} \label{sec:bhmass}

To enable comparison to CL-AGN origin theories, we calculate black hole masses from the width of H$\beta$ and the luminosity of the continuum at 5100\si{\angstrom} ($L_{5100}$). We obtain these parameters by fitting on-state SDSS spectra for each CL-AGN using the quasar spectrum fitting program \texttt{PyQSOfit} \citep[][]{guo2018,shen2019}. We describe our fit parameters in more detail in Appendix \ref{sec:pyqsofit}. We then convert the H$\beta$ FWHM and the continuum luminosity at 5100 \text{\AA} to compute a black hole mass using Equation 4 from \citet{ho2015},

\begin{equation}
    \text{Log}(M_\text{BH}) = \left[ \left( \frac{\text{FWHM(H}\beta)}{1000 \text{ km s}^{-1}} \right) \left( \frac{\lambda L_\lambda (5100 \si{\angstrom})}{10^{44} \text{ erg s}^{-1}} \right)^{0.533} \right] + a,
\end{equation}

selecting $a = 6.91 \pm0.02$, the combined fit value for both classical bulges and pseudobulges. Uncertainties in the black hole mass are dominated by the 0.35 dex intrinsic scatter in the \citet{ho2015} relation; we take this as the error on our black hole masses.

CL-AGN have intrinsically variable broad line widths, complicating black hole mass measurements. \citet{guo2024b} fit black hole mass with both the bright and dim CL-AGN spectrum and demonstrate that black hole mass measurements from H$\beta$ and $\lambda L_\lambda(5100 \text{\AA})$ are robust to CL-AGN variability. Black hole masses among our sample range from $\sim10^{7.5}-10^{8.5}$ M$_\odot$, consistent with the results of other CL-AGN surveys \citep[][]{guo2024b}.

\section{Developing a CL-AGN criterion} \label{sec:developing_criterion}

One of the aims of this work is to set out a reasonable standard for CL-AGN light curve transitions in real-time transient surveys and to determine whether these criteria can be used to follow them up. Currently, no photometric CL-AGN criterion exists, making it difficult to take advantage of surveys like the LSST to assemble large samples of CL-AGN for study. Here, we develop a photometric transition criterion using the $g$ and $r$ bands of ZTF, as these are similar to the LSST $g$ and $r$ bands \citep[][]{ztf2,ivezic2019}. To determine a reasonable transition criterion, we examine both the predicted photometric change associated with observed CL-AGN spectral transitions (Section \ref{sec:sim_phot}) and the typical variability amplitude among a sample of simulated damped random walk light curves (Section \ref{sec:drw}). We develop a criterion which maximizes purity of the selected CL-AGN photometric transitions rather than completeness; we evaluate the success of this criterion in Section \ref{sec:puritycompleteness}.

\subsection{Identifying CL-AGN transitions in the light curve} \label{sec:sim_phot}

We first examine the photometric transition properties for synthetic photometry of real CL-AGN spectra which fall in our $0.2\leq z\leq0.4$ redshift range. These transition properties can be used to inform a typical photometric transition magnitude for CL-AGN.

We simulate the expected photometric change associated with the CL-AGN transition by creating synthetic photometry using \texttt{sedpy} \citep{sedpy}. We  collect CL-AGN identified in repeat SDSS spectroscopy from 
\citet[][]{ruan2016, runnoe2016,yang2018}, and \citet{zeltyn2022}, as well as changing-look quasars (CLQs) selected with repeat SDSS spectra from \citet{lamassa2015} and \citet{green2022}.  We select restrict our sample to objects at $0.2\leq z\leq0.4$, as we do for the other samples in this paper, leaving us with 23 CL-AGN. We synthesize observed $g-$ and $r$-band magnitudes with the ZTF filterset (note that we do not deredshift spectra to the rest frame, as we wish to compare with observations of AGN which are randomly distributed in space).

We also obtain changes in color and magnitude across the CL-AGN transition for the CLQs from \citet{duffy2025}. Two of the three CL-AGN transitions identified in \citet{duffy2025} work (J1011\footnote[1]{J1011 was also identified as a repeating CL-AGN by \citet{yang2024, lyu2025, wang2025}.} and J2333) coincide with dramatic, sigmoid-like photometric changes. The third, J2336, does not. J2333 is excluded from our analysis due to its redshift ($z=0.51$). We are unable to directly compare the ZTF light curve and synthetic photometric transition for J1011, as the two SDSS spectra for this object cover the original turn-off transition identified in \citet{runnoe2016} while the ZTF light curve covers the later turn-on transition.  

\begin{figure}
    \centering
    \includegraphics[width=\linewidth]{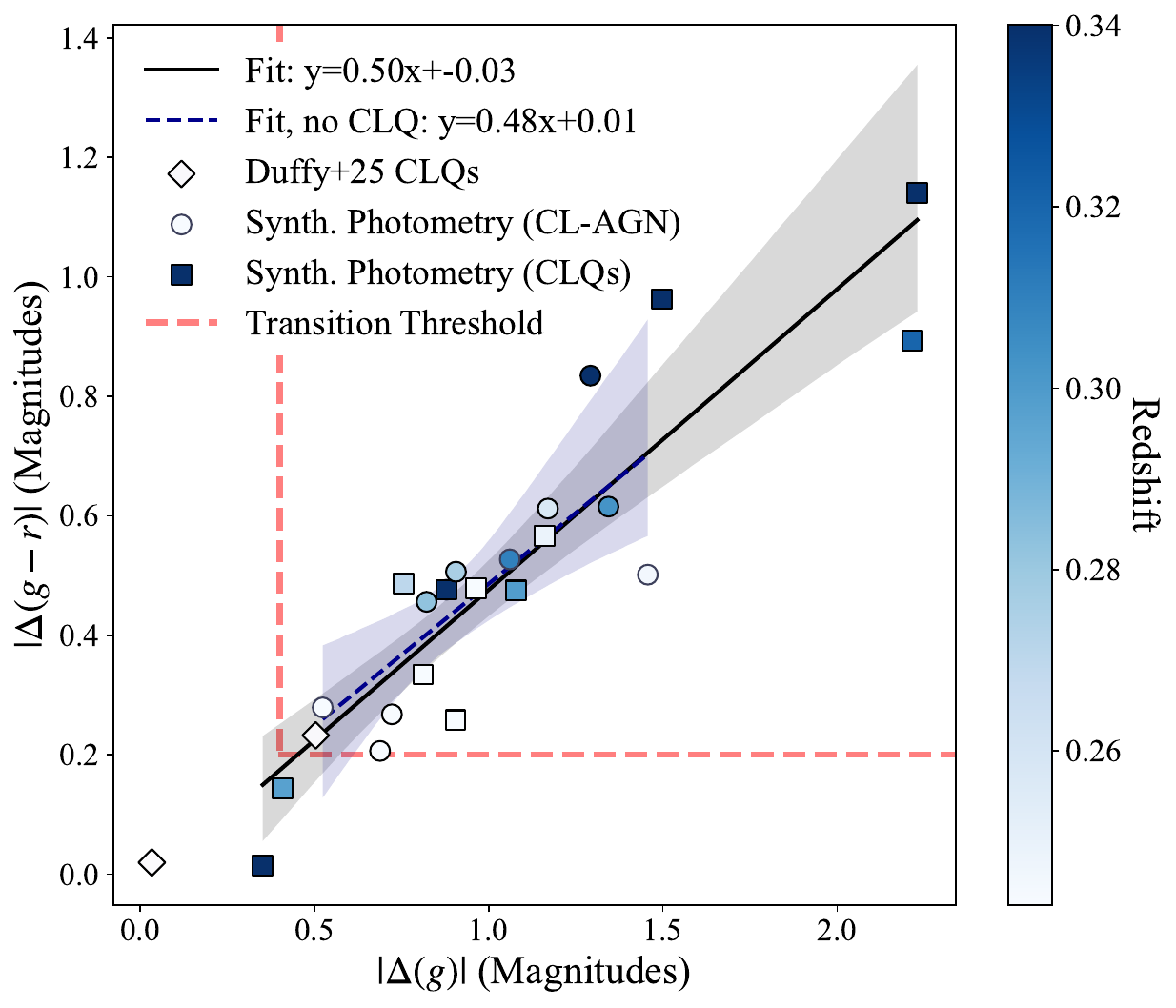}
    \caption{Predicted ZTF color and magnitude changes for CL-AGN with dual-epoch SDSS spectroscopy from \citet[][]{lamassa2015, ruan2016, runnoe2016,yang2018} and \citet[][]{zeltyn2024}; the changing-look quasars (CLQs) from \citet{green2022}; and the measured ZTF color and magnitude changes for the two changing-look quasars from \citet{duffy2025} which fall in our redshift range. The objects exhibit a range of values of $|\Delta(g)|$ and $|\Delta(g-r)|$. Based on this and the results of \citet{duffy2025}, as well as our visual identification, we select thresholds of $|\Delta(g)|>0.4$ magnitudes and $|\Delta(g-r)|>0.2$ magnitudes (red dotted lines), which would select 22 of 25 CL-AGN shown here. We find that CL-AGN transition amplitudes and colors follow the relation $|\Delta(g-r)| = (0.50 \pm 0.053)|\Delta(g)| - (0.027 \pm 0.061)$ (black line and shaded region). This relation holds within the error bars when CLQs are excluded from the linear fit (blue dotted line and shaded region).}
    \label{fig:simclagnchange}
\end{figure}

With the addition of the objects from \citet{duffy2025}, we have a total sample of 25 CL-AGN for which we can determine a typical photometric change associated with the changing-look transition. In Figure \ref{fig:simclagnchange}, we plot the absolute value of the change in $g-$ band magnitude versus $g-r$ color for our synthetic CL-AGN photometry and for the sample of \citet{duffy2025}. We perform a linear fit to only the synthetic CL-AGN photometric changes and determine that CL-AGN demonstrate a wide range of transition amplitudes and colors following the relation 
\begin{equation}
    |\Delta(g-r)| = (0.50 \pm 0.053)|\Delta(g)| - (0.027 \pm 0.061).
\end{equation}

This relationship is not significantly changed when including versus excluding CLQs in the fit. It is similar to color-magnitude relations for CL-AGN identified in the literature from multiepoch spectroscopy \citep[e.g.][]{yang2018} and light curve analysis \citep[e.g.][]{wang2025_ztf}.

\subsection{Separating CL-AGN from normal AGN variability} \label{sec:drw}

In recent years, it has been found that false periodic signals can be found in both simulated damped random walk light curves and real AGN light curves, complicating searches for quasi-periodic behavior and binary AGN in optical light curve data \citep[e.g.][]{vaughan2016,krishnan2021,davis2024,elbadry2025}. Similarly, we do not want to falsely identify CL-AGN transitions in normal AGN variability, which may randomly include prolonged increases or decreases in brightness. We therefore use simulated light curve data to determine whether we are able to separate photometric CL-AGN transitions from red noise. 

We simulate 5000 AGN light curves using the \texttt{AstroML} \citep[][]{astroml} damped random walk generator \citep[][]{kelly2009}. This function takes three arguments: the value of the structure function at infinity, which sets the variability amplitude; the value of $\tau$, or the characteristic damping timescale; and the redshift. We are interested in variability resembling that of real Seyfert AGN, so we use \texttt{qso\_fit} \citep[][]{butler2011} to measure the value of the  structure function at infinity (SF$_\text{inf}$) for the real Seyfert 1 comparison sample described in Section \ref{sec:data}. We model this distribution as a normal distribution centered at SF$_\text{inf} = 0.175$ mag with $\sigma=0.01$ mag, from which we randomly draw the values of SF$_\text{inf}$ used in mock light curve generation. We draw the value of $\tau$ randomly from a log-normal distribution centered at $\log(\tau/\rm{days}) = 2$ with a width of 1 dex, following Equation 1 of \citet{burke2021}--- 

\begin{equation} \label{eq:tau_mbh}
    \tau_\text{damping} = 107^{+11}_{-12} \text{ days } \left( \frac{M_\text{BH}}{10^8 M_\odot} \right)^{2.54^{+0.34}_{-0.35}}
\end{equation}

---for black holes in the mass range of our sample ($10^{7-9} \text{M}_\odot$; see Section \ref{sec:bhmass}). We draw the redshift randomly from the real distribution of CL-AGN redshifts in this work upsampled by a factor of 100. After generating our mock light curve, we add white noise centered at 0 mag with a width of 0.1 mag to resemble ZTF photometric noise. We again use real data to create seasonal gaps in the light curve. For each mock light curve, we sample a set of real MJD values from the Seyfert 1 sample ZTF light curves ZTF and keep only the elements of our simulated light curves which correspond to real MJDs observed in ZTF. 

Once we have generated our mock light curve with noise and observing gaps, we use the \texttt{pwlf} method described in Section \ref{sec:light_curve_fitting} to extract the values of $\Delta g$ and transition duration for our mock light curves, removing any objects for which the fit fails. We compare the recovered light curve parameters to the values for the 22 CL-AGN from \citet{ruan2016,runnoe2016,yang2018} and \citet{zeltyn2022} and the 3 CLQs from \citet[][]{duffy2025}. In Figure \ref{fig:drw_light_curve_results}, it can be seen that CL-AGN exhibit higher-amplitude changes in $g$-band magnitude than is typical for our simulated damped random walk light curves.  

To select our transition criterion, we use our best estimates of the false positive/true negative rates (the rate of damped random walk light curves above or below $|\Delta g| = 0.4$), respectively) and the true positive/false negative rates(the rate of known CL-AGN whose synthetic $g-$band magnitude changes by more or less than $|\Delta g| = 0.4$, respectively). We can identify the best criterion by maximizing the F$_1$ score \citep[see e.g.][]{hand2012}, where for $\textrm{Precision}\equiv\textrm{TP/(TP+FP)}$ and $\textrm{Recall}\equiv\textrm{TP/(TP+FN)}$,

\begin{equation}
    \textrm{F}_1 \equiv 2 \times \frac{\textrm{Precision } \times \textrm{ Recall}}{\textrm{Precision } + \textrm{ Recall}}.
\end{equation}

\begin{figure}
    \centering
    \includegraphics[width=\linewidth]{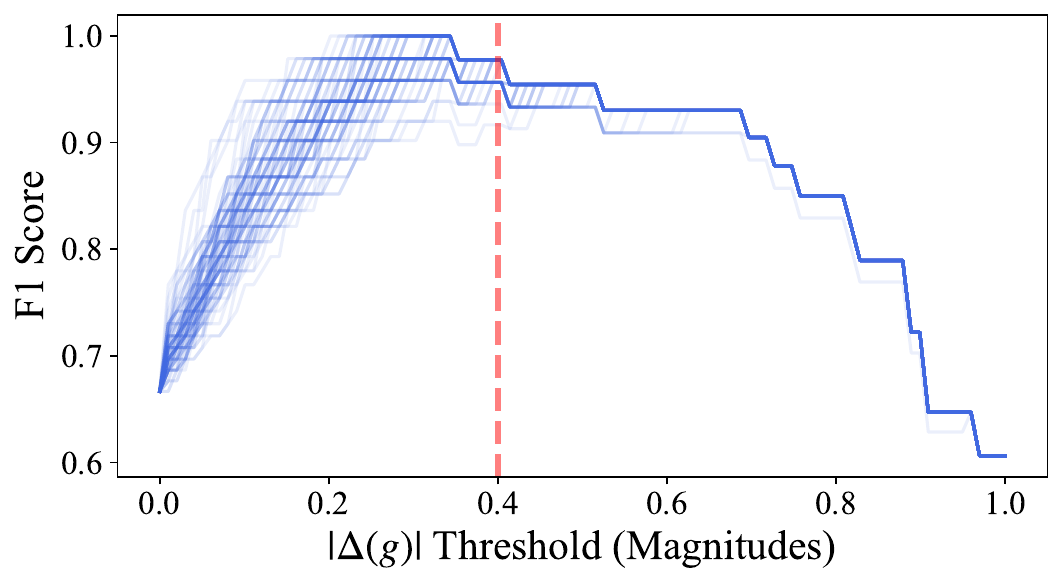}
    \caption{F$_1$ score recovered for different photometric transition criteria. An F$_1$ of one indicates perfect precision and recall. We estimate the false positive/true negative rate using our simulated damped random walk light curves and the true positive/false negative rate using simulated photometry for observed CL-AGN spectra pre- and post-transition. We account for the imbalance between our damped random walk light curves ($\sim5000$) and the number of CL-AGN with synthetic photometry (23) by randomly sampling 23 light curves 100 times to calculate the number of false positives/true negatives (blue lines). We elect to prioritize purity over completeness, leading to our selected threshold of $|\Delta g| > 0.4$ magnitudes.}
    \label{fig:f1score}
\end{figure}

\begin{figure}
    \centering
    \includegraphics[width=\linewidth]{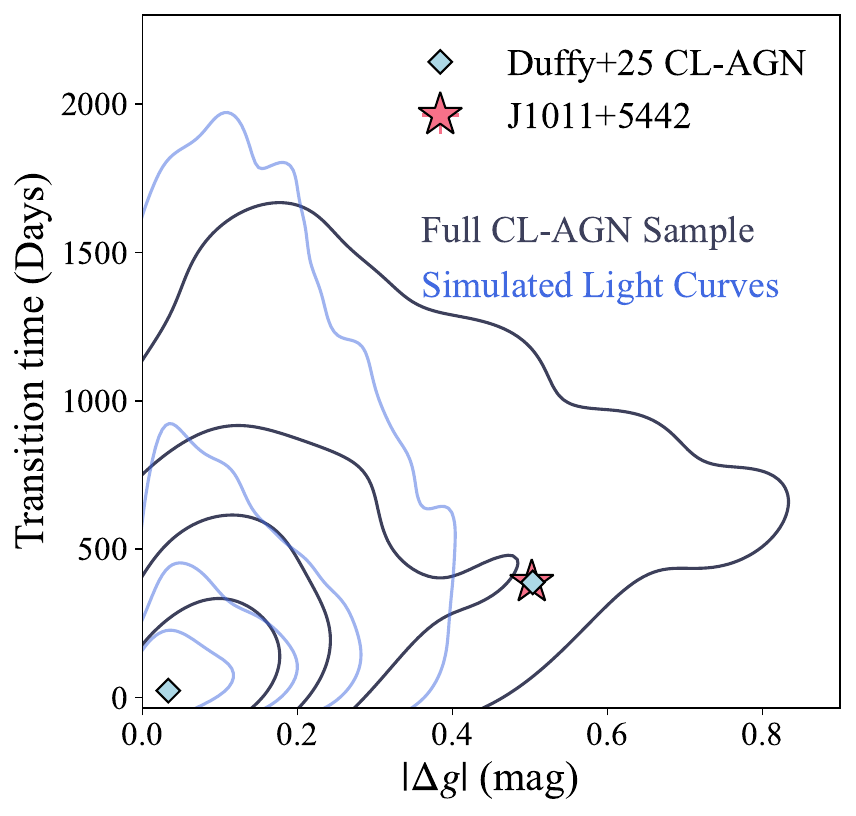}
    \caption{Comparison between CL-AGN and simulated light curve transition properties. Diamonds are repeat CL-AGN transitions identified in \citet{duffy2025}; the pink star is J1011, which is one of our photometric transition candidates. CL-AGN transition candidates occur with a higher variability amplitude than is seen for simulated damped random walk light curves.}
    \label{fig:drw_light_curve_results}
\end{figure}

An F$_1$ score equal to one indicates perfect precision and recall; we wish to maximize this score for our criterion to obtain the maximum possible precision and recall values. However, the F$_1$ score depends not on the \textit{rate} of true/false negative/positives, but rather the total number of true/false negative/positives. Here, we encounter a problem with our two tests: while we can easily simulate a large number of damped random walk light curves to estimate the number of false positives/true negatives, we are limited by the number of known CL-AGN with initial and final state spectra taken by the same instrument when we compute the true positives/false negatives. We account for this imbalance by randomly sampling 23 damped random walk light curves from our simulated data set and computing the F$_1$ score for each magnitude threshold. We repeat this process 100 times to produce the bootstrapped F$_1$ score curves shown in Figure \ref{fig:f1score}.

We select a threshold of $|\Delta g| > 0.4$ magnitudes as our photometric transition criterion (F$_1$=0.97), safely past the peak of the F$_1$ score for all bootstrapped iterations (and therefore, prioritizing purity over completeness). This threshold is somewhat conservative, as we want to be sure the photometric transitions we observe are associated with a true changing-look transition rather than normal AGN variability. We compute the rate of objects that meet this threshold, which we consider to have experienced a photometric transition. Error bars reported are $1\sigma$ binomial errors.

We find that $12^{+5.2}_{-3.7}\%$ (6/52) of CL-AGN experience a photometric transition greater than 0.4 mag in the $g$ band, while \drwrate (81/4960) of mock light curves experience a photometric transition of this magnitude (where error bars are one-sigma binomial errors). Among objects with a detected photometric transition of this magnitude, durations for known CL-AGN take a median of 430 days in the rest frame (580 days in the observer frame), with 16th and 84th percentile values of 270 and 610 days, respectively (340 and 800 observer-frame days, respectively; see further discussion in \ref{sec:clagn_in_ztf}). Uncertainties in the measured duration range from 1.1\% to 44\%, with a median error of 9.3\%. Among damped random walk light curves with variability falling above our photometric transition threshold, recovered transition durations last a median of 590 rest-frame days, with 16th and 84th percentile values of 170 and 1400 days, respectively. There is therefore significant overlap in the variability timescales relevant for normal damped random walk behavior and CL-AGN photometric transitions. This overlap informs our decision to use amplitude rather than duration of the photometric transition to develop our CL-AGN identification criteria.  

\subsection{The impact of light curve model choices}

   These results are dependent on the choices of $\tau$ and SF$_\text{inf}$ used in light curve generation. In principle, extreme damped random walk variability can mimic CL-AGN photometric transitions; if our chosen values for $\tau$ are too high or SF$_\text{inf}$ are too low, we may have an artificially low false positive rate. Among damped random walk light curves, we find the highest rate of photometric transitions in the range $1.96 \leq \text{Log}(\tau/\rm{days}) \leq 3.05$, corresponding to a black hole mass range $7.87\lesssim \text{Log}(M_\text{BH}/M_\odot) \lesssim10.6$. The black hole mass-$\tau$ relation found by \citet{burke2021} has an intrinsic scatter of $0.09^{+0.05}_{-0.04}$. While this relation implies a mass range for photometric transition candidates that overlaps with the CL-AGN in our sample, we find that we identify the highest rate of photometric transitions at $\text{Log}(\tau/\rm{days}) \approx 2.5$. This value corresponds to a black hole mass of $\text{Log}(M_\text{BH}/M_\odot) \approx 9.2$, higher than the black hole masses for CL-AGN with identified photometric transitions in our sample. CL-AGN would have to range from 6.6 to 15 $\sigma$ away from the \citet{burke2021} black hole mass-$\tau$ relation have $\text{Log}(\tau/\rm{days}) \approx 2.5$.
   
   Additionally, our results are specific to the ZTF cadence and data quality. To test whether we are picking up artificial variability due to seasonal gaps in the ZTF survey, we compare the recovered transition times among mock light curves (created with real MJD arrays from the Seyfert 1 sample ZTF data) to the time steps $dt$ between subsequent ZTF data points across the full mock light curve sample. We do not see any overrepresentation of transition durations corresponding to the seasonal gap timescale.

   Our results are also sensitive to the choice of light curve variability prescription. Recent works \citep[e.g.][]{moreno2019,yu2022,yu2025, xu2025} have modeled AGN variability as a damped harmonic oscillator, which introduces two additional degrees of freedom to model short-term variability, rather than a damped random walk process. Damped harmonic oscillators have been used to model the variability of quasars \citep[][]{yu2022, yu2025} and blazars \citep[][]{xu2025}, but not large samples of Seyfert AGN with similar bolometric luminosities to the objects in this study ($42.0 \lesssim \textrm{Log(L}_{\textrm{Bol}}/\textrm{(erg s}^{-1})) \lesssim 45.5$; see Section \ref{sec:extremevar}). Generating damped harmonic oscillator light curves at the same luminosity range as our sample would require extrapolating the observed scalings between variability amplitude and AGN luminosity, possibly producing unphysical results. The most significant differences between damped random walk and damped harmonic oscillator light curves should appear at short timescales/high frequencies \citep[for a useful illustration, see][Figure 14]{moreno2019}; longer-term variability is captured by a damped random walk process, so our results should not change significantly if we simulated damped harmonic oscillator light curves. Our estimate of the false positive rate from simulated light curves can therefore be applied to damped random walk-like variability but may not encompass the full spectrum of AGN variability, especially at short timescales.

  Indeed, when we perform our \texttt{\texttt{pwlf}} fitting to our comparison Seyfert 1 and 2 samples, we find that $3.8^{+1.4}_{-1.0}\%$ of Type 1 Seyferts (10/260) and $0.39^{+0.62}_{-0.24}\%$ (1/257) of Type 2 Seyferts experience a photometric transition with $|\Delta g| > 0.4$ mag. The rate for Type 1 Seyferts is somewhat higher than that predicted from our damped random walk light curves, indicating that some abnormal flaring behavior or CL-AGN transitions could be occurring in this sample. 
  As expected, the rate is lower for Type 2 Seyferts whose variability should be obscured by the torus. CL-AGN are therefore 3.2 times more likely than Seyfert 1s and 31 times more likely than Seyfert 2s to exhibit extreme photometric variability, even without considering changes in color.

  \subsection{Incorporating Color Evolution} \label{sec:puritycompleteness}

 We expect \drwrate of unobscured AGN to experience changes in $g$-band magnitude $>0.4$ mag due to normal AGN variability based on our simulated light curve tests. Based on the color-magnitude properties of the objects shown in Figure \ref{fig:simclagnchange}, we elect to include an additional criterion for color change of  $|\Delta(g-r)| > 0.2$. The addition of this criterion includes 22 of the 25 CL-AGN transitions shown in Figure \ref{fig:simclagnchange}, representing an estimated completeness of \completeness. This also significantly limits the number of Seyferts recovered by our transition criterion (from 10 to 3 Seyfert 1s, and from 1 to 0 Seyfert 2s). This results in a photometric transition rate of \syonerate for Seyfert 1s and \sytworate for Seyfert 2s. We require that a transition of this magnitude occur in at least 70\% of bootstrapped fits to remove cases where large uncertainties may affect the results. We now implement our criterion to identify changing-look transitions in ZTF light curves.

\section{Identifying CL-AGN in ZTF} \label{sec:clagn_in_ztf}

\begin{figure*}
    \centering
    \includegraphics[width=\linewidth]{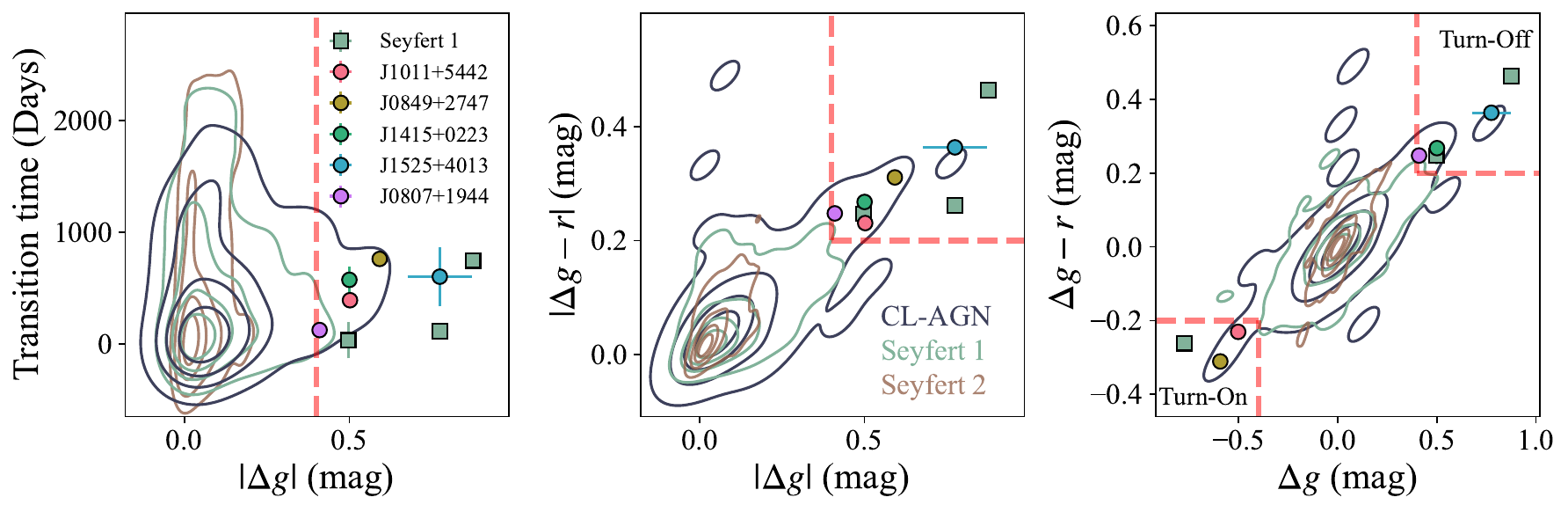} 
    \caption{Transition properties across the CL-AGN, Seyfert 1, and Seyfert 2 samples (contours), as well as for individual CL-AGN transition candidates (markers). (Left) Amplitude of the transition versus its duration. CL-AGN transition candidates have a higher variability amplitude than other AGN variability and have transition durations of a few hundred days. (Center) Magnitude of the $g$-band transition versus magnitude of the $g-r$ color transition. Our adopted threshold of $|\Delta g| > 0.4$ mag and $|\Delta (g-r)| > 0.2$ mag (red dashed line) captures only the most extreme AGN variability. (Right) Change in $g-$ band versus $g-r$ color. We find three candidate turn-on events ($\Delta g, \Delta (g-r) < 0$) and five candidate turn-off events ($\Delta g, \Delta (g-r) > 0$).}
    \label{fig:summary_props}
\end{figure*}

Here, we present the results of our light curve fitting on ZTF light curves for known CL-AGN, Seyfert 1s, and Seyfert 2s. The results of our fitting are shown in  Figure \ref{fig:summary_props}, where individual objects meeting our transition threshold are shown as colored circles (for known CL-AGN) or green squares (for Seyfert 1s; no CL-AGN photometric transitions are observed in our Seyfert 2 sample). 

\subsection{Transitions in known CL-AGN} \label{sec:known_clagn}

Five CL-AGN, or \clagnrate of CL-AGN, experience photometric transitions with $|\Delta g| \geq 0.4$ mag and $|\Delta (g-r)| \geq 0.2$ mag over the roughly six years of ZTF data in this survey (listed in Table \ref{tab:candidate_properties}). Four of the five transitions occur during the plausible transition period for that CL-AGN; in other words, we have spectra before and after the photometric event for the object with different spectral types. In the final object, we may be detecting a repeating CL-AGN event (discussed in Section \ref{sec:J0849}).

Over our sample, ZTF covers an average of $\sim15\%$ of the time between subsequent spectra used for CL-AGN identification. Assuming CL-AGN have an equal chance of transitioning at any time, and given our expected completeness of \completeness (see Section \ref{sec:puritycompleteness} and Figure \ref{fig:simclagnchange}), we would expect to observe a photometric transition in $13.2^{+0.77}_{-1.2}\%$ of known CL-AGN. This is consistent within the error bars with our detection rate.

\subsubsection{J0849: A candidate repeating CL-AGN} \label{sec:J0849}

\begin{figure*}[ht!]
    \centering
    \includegraphics[width=\linewidth]{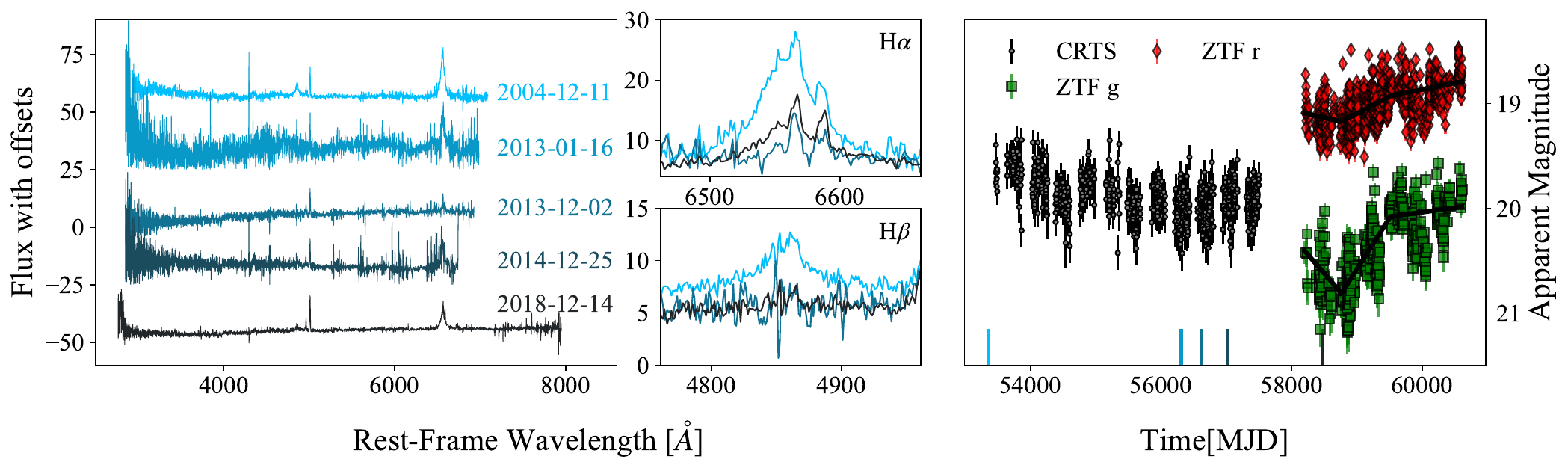}
    \caption{(Left) Spectra taken of J0849 between 2004 and 2018. Interestingly, a broad HeII feature appears at 4686\si{\angstrom} in 2013, then disappears before 2018. (Center) The broad H$\alpha$ feature (top) disappears and reappears over the course of 14 years, while broad H$\beta$ (bottom) fully disappears (spectra taken 2013-01-16 and 2014-12-25 not shown due to lack of signal to noise). (Right) Light curves from the Catalina Real Time Survey \citep[CRTS;][]{Djorgovski2011} and the Zwicky Transient Facility \citep[ZTF;][]{ztf, ztf2, ztf3}. Vertical blue lines on the x-axis correspond to the spectra in the leftmost panel. J0849 exhibits a 0.6 mag rise in the $g$-band starting in 2018 and ending in 2021.}
    \label{fig:j0849_line_changes}
\end{figure*}

For one of the five CL-AGN which meets our transition threshold, we do not have a spectrum after the photometric transition. J0849+2747 was originally identified as a turn-off CL-AGN by \citet{yang2018}. It had an archival SDSS ``QSO" spectrum at MJD 53350, but its LAMOST spectrum at MJD 57422 lacked broad-line emission. A repeat SDSS spectrum at MJD 58466 had broad H$\alpha$ again, but lacked a broad H$\beta$ component. The ZTF light curve for J0849 begins at MJD 58202, and the photometric transition we identify begins at 58970, \textit{after} the lines have re-broadened (see Figure \ref{fig:j0849_line_changes}). We take this as evidence that the LAMOST and SDSS spectra taken in 2014 and 2018 were taken during the turn-on transition, with some coincident re-brightening occurring either at low levels or before ZTF turned on.

To further confirm the changing-look transition, we use \texttt{qso\_fit}, the variability software developed in  \citet[][]{butler2011}, to measure the amount of optical variability over time. We take three time segments overlapping with each of the three spectra. We first analyze the first 1497 days of the archival CRTS light curve beginning on MJD 53470 (120 days after the archival SDSS spectrum was taken). We then select the last 1497 days of the archival CRTS light curve, which overlaps the LAMOST spectrum taken at  MJD 57422. Finally, we fit the variability characteristics of the first 1497 days of the ZTF light curve, which overlaps with the SDSS spectrum taken at MJD 58466. 

We detect QSO-like variability in the first CRTS epoch ($\sigma_{\text{QSO}}=3.4$, $ \sigma_{\text{Not QSO}}=1.1$, $\sigma_{\text{Vary}}=6.6$) as well as the final epoch of the ZTF data ($\sigma_{\text{QSO}}=10.$, $ \sigma_{\text{Not QSO}}=7.5$, $\sigma_{\text{Vary}}=42.$). In the epoch overlapping the narrow-line LAMOST spectrum, \texttt{qso\_fit} returns an ``ambiguous" classification, indictating some variability which does not necessarily follow a damped random walk ($\sigma_{\text{QSO}}=1.4$, $ \sigma_{\text{Not QSO}}=1.6$, $\sigma_{\text{Vary}}=3.8$). This is further evidence that J0849 has undergone a true repeating CL-AGN event, where both the variability and broad line emission associated with Type 1 Seyfert AGN disappeared for some amount of time between 2004 and 2018. 

With the existing data, we cannot say exactly how long the turn-off and turn-on transitions took; however, we can place an upper limit on the recurrence time (or time between spectra with broad line emission) of $\sim14$ years, with the subsequent $g$-band rise lasting 580 days in the rest frame. This is a similar timescale to that seen for other repeating CL-AGN \citep[e.g.][]{veronese2024,jana2024,wangj2024,duffy2025,dong2025,wang2025}. Further discussion of the physical implications of this recurrence timescale can be found in Section \ref{sec:clagn_origins}.

\subsection{Changing-look events in ZTF AGN} \label{sec:clagn_candidates}

We also identify photometric transitions in three Seyfert 1 AGN under the same criteria as above (listed in Table \ref{tab:candidate_properties}). This corresponds to a rate of \syonerate of Seyfert 1s experiencing a changing-look event over 6.3 years or $1.9^{+1.8}_{-1.0}\%$ over 10 years. We do not identify any photometric transitions in Seyfert 2 AGN, corresponding to a binomial one-sigma upper limit of \sytworate of Seyfert 2 AGN transitioning over 6.3 years. Again, this is expected, as $g$-band variability should be obscured by the dusty torus for most Seyfert 2s. 

This rate is consistent with the rate of damped random walk light curves which meet our criterion in $|\Delta(g)|$ (Section \ref{sec:drw}), meaning the three Seyfert 1 AGN we identify as CL-AGN candidates may simply be on the extreme end of AGN variability. This rate is also somewhat consistent with previous literature which has found a rate of $\lesssim 1\%$ of AGN change look over a decade. In particular, our rates are most consistent with those found in \citet{hon2022} ($\sim0.4\%)$) and \citet{lopeznavas2022} ($\geq0.12\%$) and higher than the rates found in \citet{yang2018} ($\sim0.006\%$) and \citet{wang2024} ($0.003-0.007\%$).  \citet{lopeznavas2022} and \citet{hon2022} only measured the rate of CL-AGN which turn on, or transition from Type 2 to Type 1; as we do not see any of these objects, it is difficult to make a direct comparison. \citet{shen2021} counted the opposite (objects which were QSO in Stripe 82 which have dimmed by at least 2 mag in g-band) and found a turn-off rate of $0.32\pm0.06\%$ over roughly sixteen years or 0.2\% over ten years. \citet{yang2018} used a combination of spectroscopic and photometric analysis in MIR, neither of which is directly comparable to our methods. \citet{wang2024} similarly used MIR variability to identify new CL-AGN. Follow-up spectra for the three Seyferts which undergo a photometric transition would allow us to determine whether they are simply experiencing normal AGN variability or whether they are CL-AGN.

\begin{table*}[]
    \centering
    \caption{Properties of candidate CL-AGN photometric transitions in our CL-AGN and Seyfert 1 samples. Durations listed are in the rest frame. (a) Originally identified as a turn-off CL-AGN in \citet{runnoe2016}; later identified as a repeat CL-AGN by \citet{yang2024,duffy2025,lyu2025} and \citet{wang2025}. (b) Sources for initial CL-AGN identification are [1] \citet{veroncetty2010}, [2] \citet{runnoe2016}, [3] \citet{yang2018}, [4] \citet{guo2024a}, and [5] \citet{zeltyn2024}.}
    \begin{tabular}{cccccccc}
        \textbf{Name} & \textbf{R.A.} & \textbf{Dec} & \textbf{Redshift} & \textbf{$\Delta (g)$} & \textbf{$\Delta (g-r)$} & \textbf{Duration} & \textbf{Source} \\
        \hline
        &&&& (mag) & (mag) & (Days) \\
        \hline
        \hline
        &&&& \textbf{Known CL-AGN}&& \\
        \hline
        J141535.46+022338.7 & 213.8977 & 2.3941 & 0.3519 & 0.50 & 0.27 & 400 & 4 \\
        J152517.57+401357.6 & 231.3232 & 40.2327 & 0.3838 & 0.77 & 0.34 & 420 & 4 \\
        J080711.72+194439.7 & 121.7988 & 19.7443 & 0.347745 & 0.42 & 0.25 & 86 & 5 \\
        J084957.78+274728.9 & 132.4908 & 27.7914 & 0.29854 & -0.59 & -0.32 & 560 & 3 \\
        J101152.98+544206.4 $^\text{a}$ & 152.9708 & 54.7018 & 0.246 & -0.50 & -0.23 & 310 & 2 \\
        \hline
        &&&& \textbf{Seyfert 1} && \\
        \hline
        J123125.89+080327.5 & 187.8579 & 8.0575 & 0.288 & 0.50 & 0.25 & 21 & 1 \\
        J021326.59+000033.4 & 33.3608 & 0.0094 & 0.294 & -0.77 & -0.26 & 86 & 1\\
        J122519.30+372053.6 & 186.3304 & 37.3481 & 0.388 & 0.88 & 0.46 & 530& 1 \\

    \end{tabular}
    
    \label{tab:candidate_properties}
\end{table*}

\subsection{Transition timescales} \label{sec:transition_timescales}

\begin{figure}
    \centering
    \includegraphics[width=\linewidth]{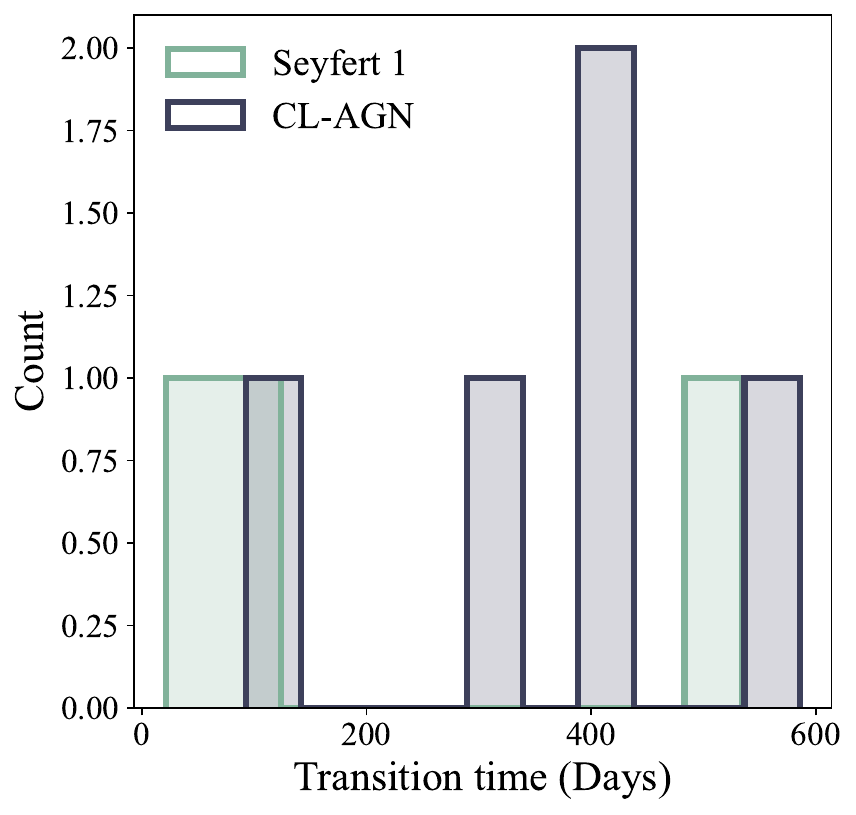}
    \caption{Rest-frame transition durations for photometric CL-AGN transitions (i.e. transitions with $\Delta (g) > 0.4$ mag and $\Delta (g-r) > 0.2$ mag). Known CL-AGN are shown in purple, while Seyfert 1s are shown in green. Photometric transitions last a median of 360 days in the rest frame ($480$ days in the observer frame), with the shortest transition being 21 days (27 days observer frame) and the longest being 560 days (760 days observer frame).}
    \label{fig:transition_time_hist}
\end{figure}

In Figure \ref{fig:transition_time_hist}, we show the distributions for transition time resulting from our bootstrapped fits for each CL-AGN transition candidate. We find rest-frame transitions ranging from 21 to 560 days, with the median time being 360 days. This is a similar timescale to those measured in \citet{duffy2025}.

\begin{figure}
    \centering
    \includegraphics[width=\linewidth]{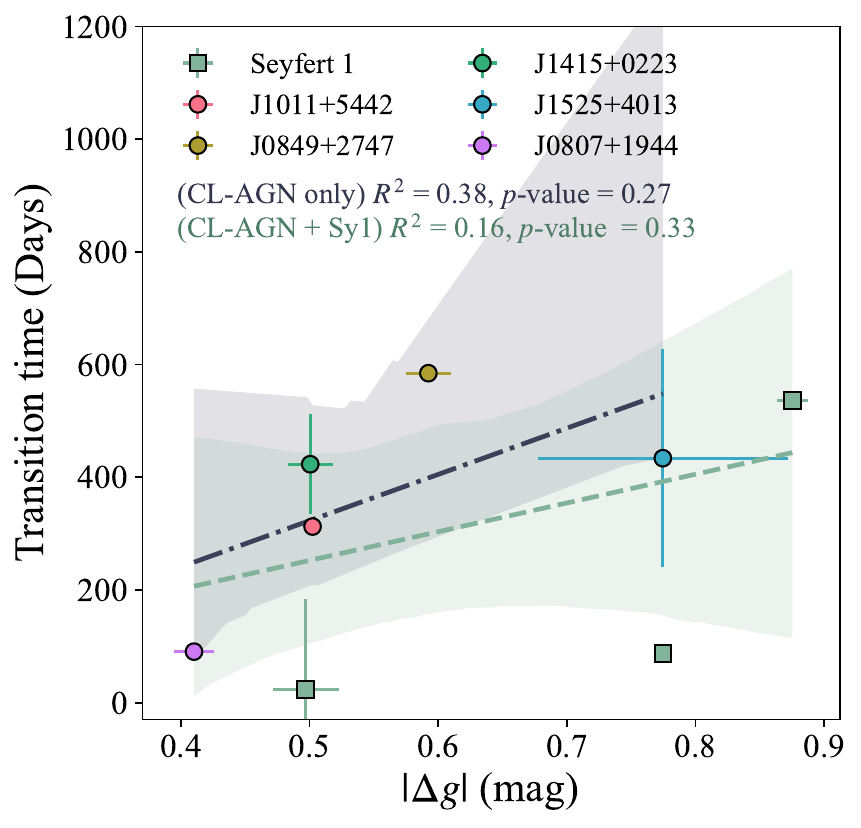}
    \caption{Change in g-band magnitude versus time for photometric transitions to occur for known CL-AGN (colored circles) and type 1 Seyferts (green squares). We perform linear regressions including only known CL-AGN (purple dash-dotted line) and including both CL-AGN and Seyfert 1s (green dashed line).  We do not detect a correlation between transition magnitude or duration at the 3$\sigma$ level for either fit.}
    \label{fig:fit_clagn_sy1}
\end{figure}

 We do not find any correlation between the amplitude of the changing-look photometric transition and the duration of that transition. In Figure \ref{fig:fit_clagn_sy1}, we show the transition time as a function of change in g-band magnitude for both known CL-AGN and candidates from the Seyfert 1 comparison sample. We attempt two linear regressions for  the sample: first, for only known CL-AGN, and second, for CL-AGN and Seyfert 1.  Regardless of which objects are included in the fit, we are unable to detect a $\geq 3 \sigma$ correlation, but we are limited by the small number of CL-AGN photometric transitions identified in this work. 

\subsubsection{Comparison to theoretical AGN timescales} \label{sec:theory_timescales}

\begin{figure}[t]
    \centering
    \includegraphics[width=\linewidth]{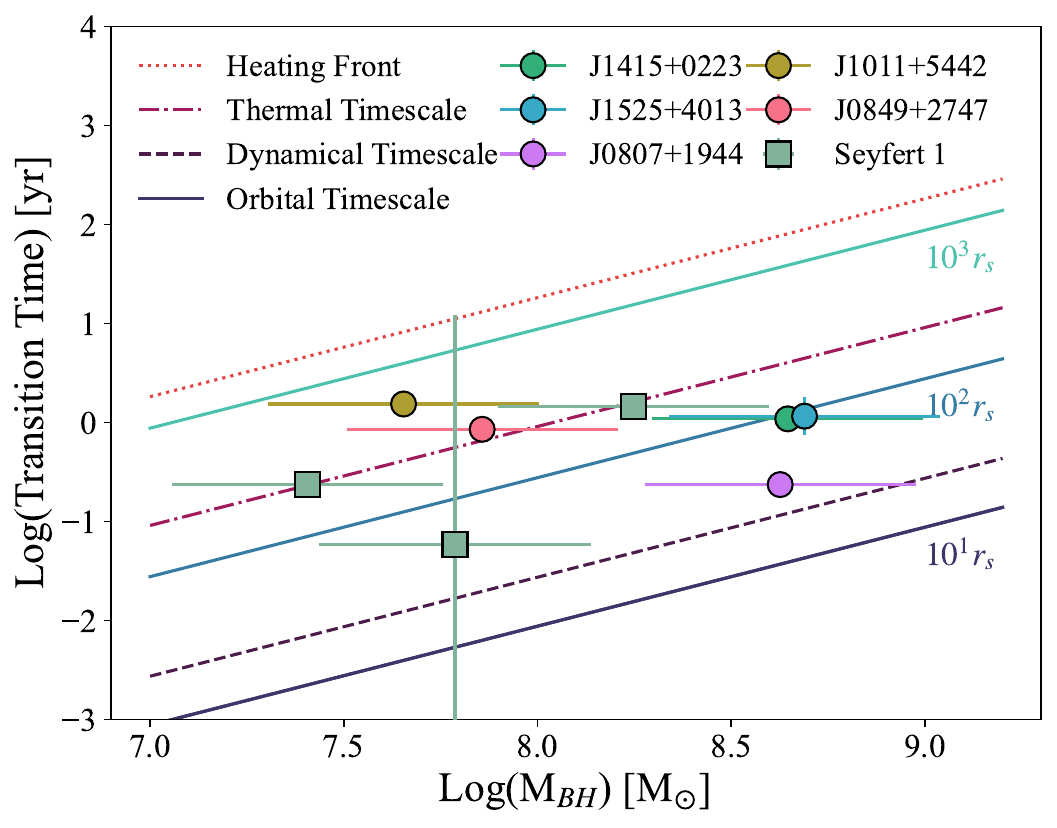}
    \caption{Rest-frame transition duration for photometric CL-AGN transitions (i.e. transitions with $\Delta (g) > 0.4$ mag and $\Delta (g-r) > 0.2$ mag) as a function of black hole mass from PyQSOfit. We compare to theoretical timescales invoked for CL-AGN transitions from \citet{stern2018,ricci2022} (dashed lines; timescales for $\textrm{R}=150\textrm{ }r_g, $ $\alpha=0.03, \textrm{ and H/R}=0.05$) and \citet{dodd2025} (solid line) and determine that CL-AGN photometric transitions are most consistent with the thermal timescale, the dynamical timescale, or the orbital timescale at $\sim10, 100, 1000 \text{ } r_s$. However, we are unable to probe timescales longer than the $\sim6$-year survey duration of ZTF (e.g. the viscous timescale). We do not detect a significant correlation between black hole mass and transition duration.}
    \label{fig:transition_props_vs_mass}
\end{figure}

As the origin of CL-AGN variability is still unknown, we next compare the timescales of our candidate changing-look transitions to the relevant timescales for AGN disks. In Figure \ref{fig:transition_props_vs_mass}, we show the duration of the changing-look transition as a function of black hole mass (see Section \ref{sec:bhmass} for fitting details) compared to predictions for the timescales for instabilities occuring at 150 $R_g$ in an AGN with $\alpha=0.03$ and $H/R = 0.05$ \citep[timescale calculations from][]{stern2018,ricci2022}. Not shown is the viscous timescale, which occurs over $10^{1-3}$ years for black holes of this mass range; this timescale is too long to be probed by the current six years of data from the ZTF survey. We also include the orbital timescales at a range of constant Schwarzchild radii; we discuss this timescale further in Section \ref{sec:tdes}.

All timescales shown in Figure \ref{fig:transition_props_vs_mass} should increase linearly with mass \citep[][]{ricci2022, dodd2021}. \citet{burke2021} includes the effect of the changing radius of the disk with black hole mass and instead predicts a correlation $t_\text{orb}, t_\text{th} \sim \text{M}_\text{BH}^{1/2}$. We do not observe any statistically significant correlation between transition duration and black hole mass among photometric transition candidates, whether we include only previously-identified CL-AGN (Pearson $r=-0.36$, $p=0.55$) or CL-AGN and Seyfert 1s (Pearson $r=0.32$, $p=0.44$). This is possibly due to the small sample of transitions we are able to identify. We explore the implications of our measured transition timescales on CL-AGN further in Section \ref{sec:clagn_origins}.

\section{Implications for CL-AGN origins} \label{sec:clagn_origins}

\subsection{Changing-obscuration}

CL-AGN may be caused by a dust cloud obscuring the BLR (or, in the case of X-ray CL-AGN, the corona). X-ray changing-obscuration events can occur on timescales as short as hours \citep[e.g.][]{elvis2004, Risaliti2007, risaliti2011,sanfrutos2013}, corresponding to the time required for a cloud of obscuring material to move in front of the X-ray corona. Clouds which obscure the BLR are more likely to be part of the dusty torus, and the relevant timescale for these events is the dynamical time in the torus. 

For an AGN with $\lambda L_\lambda(5100 \rm{\AA}) \sim 10^{44}$ erg s$^{-1}$, the radius of the BLR is $\sim30$ light-days and the radius of the torus is $\sim0.1$ parsec \citep[][]{kaspi2000,koshida2014}. Assuming a cloud in the torus must fully obscure the BLR, the crossing time for a cloud to fully obscure the BLR would be 

\begin{equation}
    \Delta t_c = \sqrt{\frac{R_c D_s^2}{G M_{BH}}} \simeq 1.3 \times 10^5 \left(\frac{\text{M}_{\text{BH}}}{\text{ M}_{\odot}}\right)^{-1/2} \text{ years.}
\end{equation}

For a $10^8$ M$_{\odot}$ black hole, this corresponds to a crossing time of around 13 years, longer than the CL-AGN transition timescale we are able to measure in this work. The long timescale required to fully obscure the BLR suggests that the photometric CL-AGN transition we observe does not occur due to obscuration by clouds in the torus. 

On the other hand, \citet{zeltyn2022} identified two possible changing-obscuration events in SDSS J162829.17+432948.5 ($z = 0.2603$) within a couple of years. The first was a long-term dimming event from 2020--2021, which was followed by a re-brightening event over $\lesssim2$ months with spectral changes consistent with a change in dust obscuration. To explain the rapid change in obscuration, they suggested that a change in the UV continuum could sublimate dust in the line of sight on a timescale slightly longer than the light crossing time at the dust sublimation radius, a scenario which was also explored by \citet{lamassa2015} to explain the behavior of the changing-look quasar SDSS J015957.64+003310.5. If we again assume a BLR with a radius of 30 light-days, this is shorter than the majority of the photometric transitions we observe. Allowing extra time for dust to sublimate or re-form (\citet{zeltyn2022} estimates $\sim1$ month for $n\simeq10^{9}$ cm$^{-3}$, with $t \propto n^{-1}$ citing \citet{draine2009}), this explanation is plausible for the objects in our sample.

However, \citet{zeltyn2022} finds that WISE data taken during the initial dimming of J162829.17+432948.5 suggests an intrinsic change in the accretion state of the disk, as mid-IR emission 1) should be coming mostly from the dusty torus, and therefore out of the line of sight and 2) less affected by dust reddening. In our sample, we lack sufficient WISE data to monitor the change in reddening through the optical photometric transition. Thus, we are unable to fully prove or disprove a changing-obscuration explanation for our sample, as multiwavelength and/or polarization data is necessary to fully confirm or rule out a change in obscuration \citep[see also][]{hutsemekers2017,sheng2017, stern2018,yang2018}.

\subsection{Are CL-AGN the ``extreme end" of normal AGN variability?} \label{sec:extremevar}

 \citet{guo2024b} proposed some CL-AGN may just be experiencing normal AGN variability. CL-AGN occur mainly on the lower end of the AGN luminosity distribution \citep[][]{macleod2019,green2022,wangj2024,guo2024b}. Luminosity is anticorrelated with variability \citep[][]{wilhite2008, zuo2012, simm2016, Chanchaiworawit2024}, so it would be reasonable that (at least some) CL-AGN are simply on the high-variability tail of AGN behavior. This is supported by the findings that CL-AGN preferentially occur at low Eddington ratios, especially at $-2 \leq \text{Log}(\lambda_\text{Edd}) \leq -1$ \citep[e.g.][]{zeltyn2024,guo2024b,jana2024}, though this may be partly due to the relatively small luminosity changes necessary to impact broad line emission in an already-dim source \citep[][]{komossa2024,lu2025}. 

\begin{figure}
     \centering
     \includegraphics[width=\linewidth]{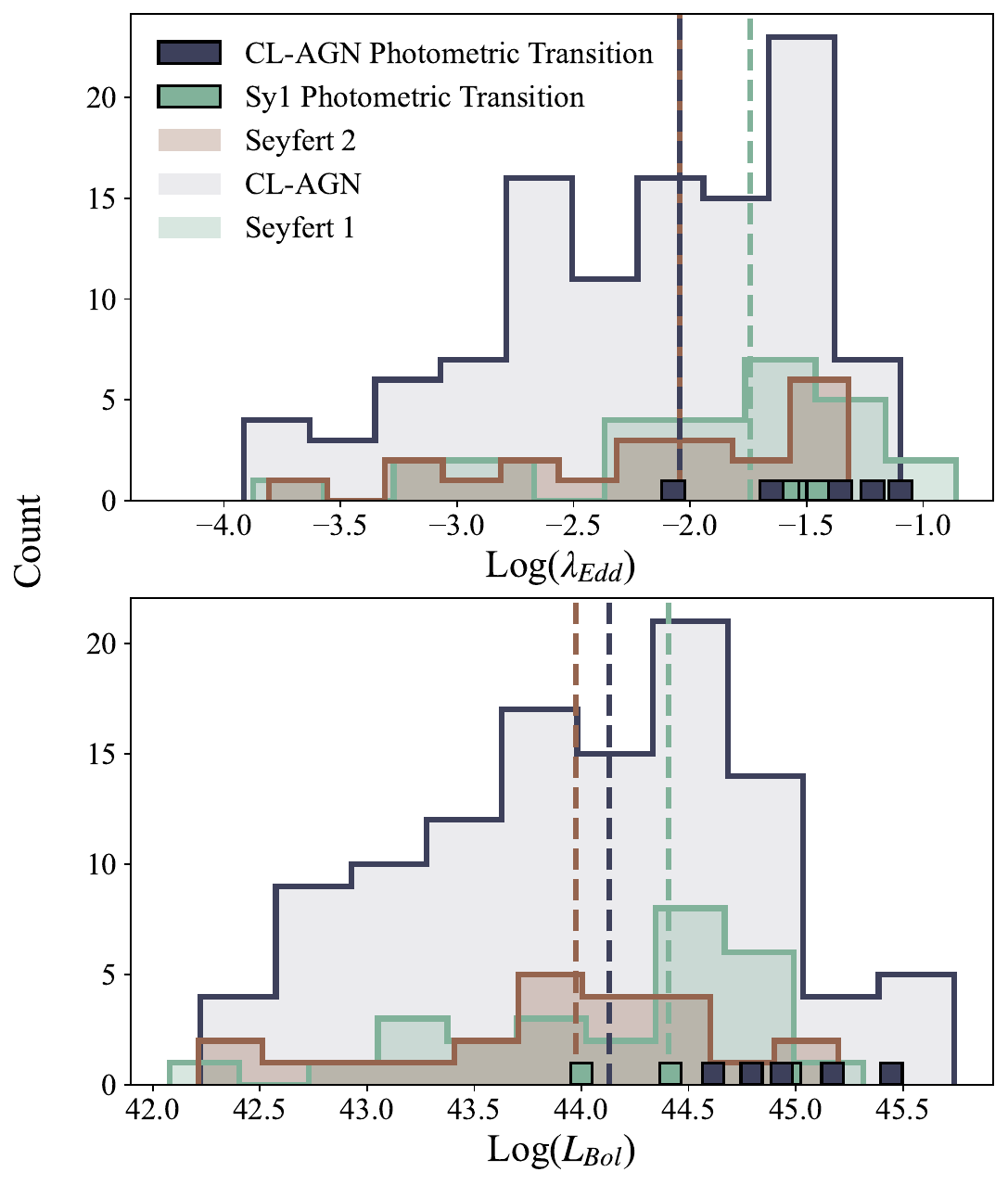}
     \caption{Distribution of Eddington ratios (top) and bolometric luminosities (bottom) for the CL-AGN photometric transition candidates (bar graph) and all objects studied in this work (histograms). Median values for each sample are shown with dashed vertical lines. Photometric transitions occur mainly in objects with an Eddington ratio $-2 \lesssim \text{Log}(\lambda_\text{Edd}) \lesssim -1$ (measured from the on-state spectrum). A Kolmogorov-Smirnov test finds photometric transition candidates have marginally statistically distinct Eddington ratios from Seyfert 2s ($p=0.051$) and Seyfert 1s ($p=0.035$), though CL-AGN overall are not ($p=0.17$ for Seyfert 2s and $p=0.22$ for Seyfert 1s).}
     \label{fig:lumdist}
 \end{figure}

To compute the Eddington ratio ($\lambda_\text{Edd} \equiv L_\text{Bol}/L_\text{Edd}$), we use \texttt{PyQSOFit} to measure $\lambda L_\lambda(5100\text{\AA})$ for the SDSS spectrum of each object. Following the analysis of \citet{zeltyn2025}, we apply a bolometric correction of 9.26 \citep[][]{richards2006,shen2008}. We use the H($\beta$) black hole masses used throughout this work and Equation 3 from \citet{zeltyn2025},

\begin{equation}
    \lambda_\text{Edd} \equiv \frac{L_\text{Bol}}{1.5 \times 10^{38} \times (\text{M}_\text{BH}/\text{M}_\odot)  \text{ erg } \text{s}^{-1}} ,
\end{equation}

 to compute the Eddington ratio for the CL-AGN in our sample as well as for a redshift-matched subset of the Seyfert 1s and Seyfert 2s from our comparison sample (Figure \ref{fig:lumdist}).  We do find that we are more likely to identify photometric CL-AGN transitions in CL-AGN at $-2 \leq \text{Log}(\lambda_\text{Edd}) \leq -1$ ($p=4.05 \times 10^{-3}$ under a Kolmogorov-Smirnov test). However, we do not find evidence that our CL-AGN sample is accreting at a systematically different Eddington ratio than comparison Seyfert 1 and 2 AGN ($p=0.22$ and $p=0.17$, respectively). A KS test performed for bolometric luminosity rather than Eddington ratio again finds no statistically significant difference between our CL-AGN sample and Seyfert 1s ($p=0.26$) or 2s ($p=0.53$), though again objects with a detected photometric transition are borderline statistically distinct from CL-AGN ($p=0.010$) and Seyfert 1s ($p=0.070$) and are statistically distinct from Seyfert 2s ($p=6.04 \times 10^{-3}$). The latter may have more to do with the greater obscuration for Seyfert 2s, which also prevents us from observing a photometric transition among that population.

If the observed Eddington preference for CL-AGN were due solely to increased variability as AGN luminosity decreases, we would expect to find that CL-AGN have systematically lower Eddington ratios and bolometric luminosities than other Seyferts. We do not find this to be the case. The slight clustering of photometric transitions at higher Eddington ratios and in the higher half of bolometric luminosities among our sample may reflect a selection bias in the more luminous CL-AGN are less host-dominated, and thus their transitions' variability are easier to detect in optical photometry. However, we caution that the Eddington ratio has been observed to change over the course of the CL-AGN transition \citep[e.g.][]{noda2018,duffy2025b,guo2024b,jana2024}, while our single-epoch Eddington ratio measurements are measured for the on state spectra only. These measurements should therefore be taken as a maximum luminosity and Eddington ratio for our CL-AGN sample.

\subsection{Changes to the accretion disk} \label{sec:diskinstabilities}

If CL-AGN are not caused by changes in line-of-sight obscuration or a normal extension of AGN variability, it is reasonable to ask which mechanisms could cause the intrinsic accretion rate of the AGN to change dramatically enough to cause the observed spectral transition. Here, we discuss the relevant timescales and photometric appearances predicted for changing-state transitions of various kinds.

\subsubsection{Tidal disruption events} \label{sec:tdes}

Tidal disruption events (TDEs) inside AGN disks have been invoked to explain the sudden accretion rate change seen in CL-AGN. The timescale for TDEs around quiescent black holes is approximately one year, with $\sim$month-long rise and a characteristic $t^{-5/3}$ mass fallback rate \citep[][]{rees1988, zabludoff2021}. \citet{guo2024b} proposed at least some CL-AGN \citep[e.g.][]{shappee2014,gezari2017,Frederick2019} are transient flares or outbursts triggered by a sudden accretion event like a TDE. AGN are usually excluded from TDE searches as their intrinsic variability and occasional flares make it difficult to distinguish normal AGN behavior from TDEs in an AGN disk. However, it is extremely likely that TDEs occur in AGN disks, and some theories posit that AGN should increase the TDE rate at least temporarily \citep[][]{wangm2024,kaur2025}.

We find that CL-AGN transitions occur on longer timescales than TDEs around quiescent black holes. The  median duration of a changing-look transition across our sample is 360 days. The most rapid transition in our sample occurs over 21 days, consistent with the rise time of a TDE; however, this is a turn-off transition in one of our Seyfert 1 CL-AGN candidates, and it may also be an example of normal AGN variability (see light curve in Appendix \ref{sec:light_curves}).  

The longer transitions we observe in CL-AGN do not necessarily rule out TDEs in AGN disks. Simulations of TDEs occuring in AGN disks predict a variety of timescales and light curve shapes. \citet{chan2019} found that the fallback time for a TDE in an AGN should be 4 days, with the diffusion cooling time starting at 21 days before slowing down and the mass disk being accreted over 60 years. Under this model, it would be reasonable to have a sigmoid-shaped light curve on the six-year timescale probed by ZTF. \citet{ryu2024} found that in sufficiently dense AGN disk, a TDE will be muffled by interactions with the disk, suppressing the TDE light curve. However, passage through inner disk and fallback onto the disk will puff up the disk, leading to an increased accretion rate and state changes similar to those invoked for CL-AGN. They predicted a thermal timescale of $\sim$ days for this initial temporary increase in accretion rate, then a viscous timescale of $\sim$6 years for the puffed-up inner disk to be fully accreted. These timescales are not consistent with our results, but their simulations don't yet take into account radiation pressure, which can puff up the accretion disk and decrease the thermal and viscous timescales. Recent work incorporating the effects of inclination angle and apsidal precession found the flare timescale for a TDE around an AGN to be $\sim20$ days in the case of a $10^6 \text{M}_\odot$ black hole \citep[][]{zhang2025}. This is a factor of 10 lower than our results for $\sim10^{8} \text{M}_\odot$ black holes, or a scaling factor of $\sim\text{M}_\text{BH}^{1/2}$, consistent with scaling for the thermal timescale \citep[][]{burke2021}. In this scenario, the timescales we measure could be associated with a low-inclination, prograde TDE in a pre-existing AGN disk.

Two examples of candidate TDEs in active AGN disks may shed some light on this origin for CL-AGN. \citet{ricci2020} suggested that a TDE could disrupt the X-ray corona and explain the light curve seen for 1ES 1927+654, which changed look over $\sim3$ months and gained broad line emission over about six weeks \citep[][]{Trakhtenbrot2019}. This CL-AGN does not have a similar light curve or transition timescale to the objects in \citet{duffy2025} or to our sample, indicating it may be an outlier object (perhaps a TDE in a low-density AGN disk, e.g. \citealt{ryu2024}, or a near-perpendicular disruption relative to disk, e.g. \citealt{zhang2025}). \citet{blanchard2017} identified a flare in a known narrow-line Seyfert 1 galaxy which they attributed to a TDE in an AGN disk. PS16dtm experienced a $\sim50$ day rise followed by a $\sim100$ day plateau, after which the UV magnitude began to decline. PS16dtm's color remained relatively steady throughout the rise and plateau, and it showed little spectral evolution. This rapid rise is similar to the timescale of the rise for 1ES 1927+654 and is again much shorter than the rise seen in all but one of the CL-AGN transition candidates we identify in this work. However, \citet{blanchard2017} propose that the TDE which caused the flare occurred off-axis relative to the existing AGN disk, forming its own elliptical accretion disk which blocked PS16dtm's X-ray accretion (seen in archival data of the host galaxy). This may indicate that both 1ES 1927+654 and PS16dtm resemble a TDE + AGN because the two accretion processes are distinct, while CL-AGN may be caused by a deposition of material by the TDE onto the AGN disk, increasing the accretion rate through the AGN disk. 

\subsubsection{Thermal-viscous disk instability}

The thermal-viscous instability is a limit-cycle instability thought to cause the hard-soft state change observed in X-ray binaries and cataclysmic variables \citep[see review by][]{lasota2001}. This instability occurs when the accretion disk crosses $10^4$ K, the temperature at which hydrogen ionizes. Crossing this temperature changes the opacity of the disk, which changes its viscosity. As the viscosity changes, so does the accretion rate. This transition occurs around an Eddington ratio $\text{log}(\lambda_\text{Edd}) \sim -2$ \citep[][]{Maccarone2003,done2007,yang2015}, similar to the Eddington ratio at which many CL-AGN transitions occur \citep[][]{guolo2021,panda2024,jana2024,guo2024b}.

This has led to the theory that CL-AGN are undergoing the same disk instability mechanism as X-ray binaries. \citet{hameury2009} first constructed a framework for the thermal-viscous disk instability in AGN which should occur on timescales of $\sim1000-10^5$ years, too long to explain normal AGN variability or flares. \citet{noda2018} argued that the sound speed in AGN should be faster due to impact from radiation and magnetic pressure, reducing the viscous timescale and possibly explaining the changing-look transition in Mrk 1018. \citet{wu2023} similarly argued for an increased sound speed from magnetic pressure and outflows, reproducing timescales as low as $10^{-2}$ years for $\text{M}_\text{BH} \sim 10^6 \text{M}_\odot$. (See their Figure 3 for a mass-dependent timescale versus our Figure \ref{fig:transition_props_vs_mass}.) \citet{veronese2024} cited the appearance of a radio jet in Mrk 1018 as evidence in favor of this mechanism for CL-AGN. Jet launching has also been observed in 1ES 1927+654 \citep[][]{laha2022, laha2025}, though young radio jets are not more common among CL-AGN than among other kinds of AGN \citep[][]{wuye2023,birmingham2025}. Spectral modeling for the CLQs studied in \citet{duffy2025} demonstrated that the low state SED for CLQs can be qualitatively similar to ADAF + thin disk models, further supporting a transition in disk state \citep[][]{duffy2025b}.

Our measured CL-AGN transition durations are far too short to be explained by the viscous timescale in a standard AGN accretion disk, but they are consistent with the timescales predicted by \citet{wu2023} for a magnetically-supported accretion disk with outflows. Importantly, as this is a limit-cycle mechanism, this theory implies that many or most CL-AGN should repeatedly switch between the on and off state. \citet{wangj2024} compiled eleven repeating CL-AGN which had been identified as of 2024; \citet{wang2025} identified eight repeating CL-AGN, one of which (J1011) overlaps with the three identified in \citet{duffy2025}; \citet{dong2025} recently identified a further 19. In this work, we have identified another repeating CL-AGN candidate, J0849+2747. The most rapid repeating CL-AGN is NGC 7603, which repeated within $1-2$ years \citep[][]{tohline1976,Kollatschny2000,wangj2024}; the slowest was Mrk 1018, which took between $25-36$ years to repeat \citep[][]{osterbrock1981,cohen1986,mcelroy2016,wangj2024}. The average CL-AGN in \citet{wangj2024} took $6.4-16.5$ years to repeat; this means repeated spectroscopic monitoring over the course of a decade or more is necessary to identify repeating CL-AGN. While the current fraction of repeating CL-AGN is low (41 of the $\sim100$s of CL-AGN which have been identified to date), the number of CL-AGN with a sufficient baseline to identify a repeating CL-AGN event is also low. It is possible that future monitoring campaigns will find that many or even most CL-AGN repeat, possibly with some periodicity. If so, this will be a strong piece of evidence in favor of the thermal-viscous disk instability as a physical mechanism for CL-AGN.

\subsubsection{Radiation pressure instability}

\citet{sniegowska2020} described radiation pressure instability as a possible driver of CL-AGN. Under this model, CL-AGN occur due to radiation pressure instability between the stable outer disk and the advection-dominated accretion flow. \citet{sniegowska2023} explored timescales for this instability around black holes with masses of 10, $10^5$, and $10^7$ M$_\odot$ and determined outburst behavior is primarily determined by the magnetic field and the outer radius of the disk. Decreasing the radius of the disk decreases the timescale for the instability-driven outburst. Magnetic fields begin by increasing the period of the outburst, then decreasing it after some critical value. \citet{sniegowska2023} found that the period of this instability can explain quasi-periodic eruptions \citep[QPEs; ][]{miniutti2019} in a TDE-sized disk. To explain CL-AGN, the accretion disk would either need to be smaller than expected or to contain a gap, possibly from a secondary black hole. 

Limit cycle durations for the radiation pressure instability can reach a decade when they occur in a very small accretion disk ($R_{out} \sim 100 R_g$). Measurements of the period between repeating CL-AGN transitions range from a few years to over a decade \citep[e.g.][]{wangj2024, duffy2025, dong2025}, but these studies are limited by the small number of mostly-nearby AGN which have been monitored for over a decade. As with the thermal-viscous instability, long-term photometric and spectroscopic monitoring of large samples of known AGN with LSST and Roman will be necessary to determine how many CL-AGN exhibit repeat transitions and on what timescales.

\subsubsection{Propagation of a heating/cooling front}

\citet{ross2018} proposed a scenario in which changing-look quasars are triggered by the propagation of a heating or cooling front through the accretion disk. The timescale for this to occur is given by

\begin{equation}
    t_\text{front} \sim 20 \text{ yrs} \left( \frac{h/R}{0.05} \right)^{-1} \left( \frac{\alpha}{0.03} \right)^{-1} \left(\frac{M_\text{BH}}{10^8 M_\odot}\right) \left( \frac{R}{150 r_s} \right)^{3/2},
\end{equation}

where the viscosity and scale height of the disk moderate the speed at which the front can propagate \citep[][equation 7]{stern2018}. \citet{ross2018} proposed that changes at the innermost stable circular orbit, or ISCO (where $r_\text{ISCO} \equiv 6r_s$) could deflate the inner disk, sending a cooling front outward. At later times a heating front would propagate inward, re-inflating the disk to its original state. This scenario could explain the $\sim20$ year dimming and re-brightening of J110057.7-005304.5 they observed.

\citet{stern2018} observed a dimming event of 1.8 magnitudes in the $g$-band in the source WISEJ1052+1519 over $\sim10$ years. They similarly proposed that an instability at or near the ISCO, perhaps due to torque from material plunging over the ISCO, could cause temperature and thickness changes in the inner disk, allowing a heating or cooling front to propagate more quickly than expected in a thin-disk approximation. 

In principle, this mechanism could also explain the observed dimming and re-brightening of the objects from \citet{duffy2025}. However, as was pointed out in \citet{duffy2025}, the model presented in \citet{stern2018} predicts a difference a shorter turn-off timescale than turn-on, which was not observed for the three CLQs in the \citet{duffy2025} sample. We do not observe a difference in the turn-on and turn-off timescales for our photometric transition candidates, but this may be due to the small number of objects with a candidate photometric transition or because the turn-on and turn-off timescales should depend on properties which vary across the sample (e.g. black hole mass, disk viscosity and scale height). Further photometric monitoring for dimming and re-brightening events will be crucial to test this theory.

Our observed transition durations are a factor of 10 shorter than the durations observed in \citet{ross2018} and \citet{stern2018}. If the photometric transitions in this study are caused by the propagation of a heating or cooling front, the disk would need to be inflated by a factor of $\sim10$ to explain the observed transition duration. Additionally, \citet{noda2018} stated that the propagation of a thermal front should be accompanied by dramatic photometric change but not a spectral change, unless the disk also evaporates or condenses, which they invoked as the explanation for the repeating CL-AGN behavior in Mrk 1018. \citet{graham2020} found that the combination of extreme photometric and spectroscopic variability in their sample on $\sim$years timescales supports the combination of a heating/cooling front and a disk transition to produce their observed sample of changing-state quasars, which vary on timescales comparable to those observed in our study.

\subsubsection{Disk tearing}

Disks can warp due to Lens-Thirring precession around a spinning black hole; this warp can cause an accretion disk to break into rings \citep[e.g.][]{raj2021_sim}. These rings can have shocks at boundaries, slowing down rotation and forming an inner elliptical accretion disk accreted on the viscous timescale. \citet{raj2021_timescales} proposes disk tearing as a potential cause of CL-AGN and explores relevant timescales for luminosity changes due to disk tearing and disk precession. Disk tearing can deliver material to small radii where viscous accretion occurs on short timescales. At small radii, radiation pressure decreases timescale further, leading to accretion timescales as short as weeks or as long as years. 

\citet{liska2023} simulated a tearing event in a thin accretion disk and found that tearing events could lead to bursts of accretion that last days--a month for black holes in the mass range $10^{7.5-8.5} M_\odot$. Here, outburst magnitude is determined by both changes in the accretion rate due to disk tearing and changes in the radiative efficiency of the disk. These outbursts would also be modulated by precession of both the outer and torn inner disk, adding additional signal at higher frequency. Recently, \citet{kaaz2025} produced synthetic observations of the disk from \citet{liska2023} and found that the outbursts could lead to order-of-magnitude changes in the bolometric luminosity and width of H$\alpha$ and H$\beta$ on timescales of $\sim100$ days for a $10^8 M_\odot$ black hole, the same order of magnitude as our results for black holes of similar masses.

Here again, long-term monitoring will be crucial for confirming disk tearing as a mechanism to produce CL-AGN. The accretion outbursts modeled in \citet{liska2023} and \citet{kaaz2025} occur semi-periodically, with \citet{kaaz2025} finding multiple peaks in the light curve over 600 days. Additionally, the exact duration of the outburst is also determined by the radius at which the disk tears, which depends on the disk parameters. Thus, better constraints on these parameters for the accretion disks in our sample would be useful to determine whether the observed photometric changes could be explained by disk tearing.

\subsubsection{Spiral shocks}

\citet{wang2020} proposed that CL-AGN could be caused by a close supermassive black hole binary inside a circumbinary disk, supported by observations that CL-AGN may occur more commonly in hosts which have undergone recent mergers (\citealt{charlton2019}, but see also \citealt{dodd2021}, \citealt{liu2021}, and \citealt{agrawal2025}). Each SMBH peels gas off the circumbinary disk, forming its own accretion disk. Torques between the SMBHs can induce spiral shocks in the disk, whose propagation should alter the accretion rate. Recently, \citet{dodd2025} developed an analytic model for perturbation of the accretion disk by a companion object to the central SMBH as a theory for turn-on CL-AGN. This object could in principle be a star or another black hole; \citet{dodd2025} argued that in the case of CL-AGN, the timescales and flare amplitudes involved suggest a massive or supermassive black hole companion rather than a star or stellar-mass compact object. Under their theory, a companion object moving through the AGN disk triggers a spiral density wave, temporarily increasing the accretion rate. The flare's duration would be at minimum the orbital timescale \citep[][equation 2]{dodd2025}, 

\begin{equation}
    t_\text{orb} = 2 \pi \sqrt{\frac{r^3}{GM_\text{BH}}}\text{ .}
\end{equation}

In Figure \ref{fig:transition_props_vs_mass}, we show the orbital timescale as a function of black hole mass at 10, 100, and 1000 $r_s$. The transition durations measured in this work are consistent with a flare produced by spiral density waves triggered by disk perturbation within this range. 

\subsection{Molecular gas transition} \label{sec:gas_transition}

Finally, it has been proposed that CL-AGN are caused by a change in the available molecular gas fuel to the accretion disk from the host galaxy. \citet{liu2021} and \citet{wang2023} suggested that CL-AGN may occur in the transition between ``feast" and ``famine" fueling to an AGN, where ``feast" fueling occurs when an AGN has a steady supply of cold molecular gas to the disk and ``famine" fueling occurs when the AGN instead accretes material shed by aging stars in the nuclear star cluster \citep[][]{kauffmann2009}. Some studies have found that CL-AGN reside primarily in Green Valley host galaxies \citep[][]{liu2021,dodd2021} or galaxies with a large number of intermediate-age stars \citep[][]{jin2021}, though other studies find similar stellar populations between CL-AGN and non-CL-AGN host galaxies \citep[e.g.][]{charlton2019, yu2020, verrico2025, zeltyn2025}. \citet{liu2021} and \citet{wang2023} proposed that the same processes shutting off gas supply to these galaxies and causing quenching are cutting off the supply of gas to the nucleus, causing CL-AGN to exhibit unstable accretion as their fuel source is depleted.

Similarly, \citet{shen2021} argued that changing-look transitions are the beginning or end of the 10$^{4-5}$ year AGN duty cycle \citep[][]{Schawinski2015}. Under this model, some small percentage of AGN should always be transitioning from the high- to low-accretion state, supported by the low rate of CL transitions in a sample of Stripe 82 quasars \citep[$0.32\pm0.06\%$ over roughly sixteen years;][]{shen2021}. Modeling changing-look events as the end of the quasar lifetime, \citet{shen2021} argued for a quasar lifetime of 10$^{2-4}$ years, consistent with the viscous infall timescale from $R\sim 100 R_g$ for a 10$^{8-9} M_\odot$ black hole. The decreasing fuel availability at small radii may also lead to instabilities which mimic faster-timescale changing-look behavior. \citet{guo2024b} similarly suggested that some CL-AGN are intrinsically fading at the end of the AGN duty cycle (as opposed to objects which exhibit extreme variability or which are experiencing a transient flaring event). They point to a subset of objects which have faded and lost broad-line emission over several decades as evidence of this phenomenon \citep[longer than the $\sim10$ year DRW characteristic timescale; see][]{kozlowski2016,stone2022}.

\citet{shen2021} used the formula $t_\text{ep} = \Delta t /f_\text{TO}$ to calculate the quasar episode lifetime, where $\Delta t$ is the baseline for observations and $f_\text{TO}$ is the fraction of quasars which turn off in that time. As stated in Section \ref{sec:clagn_candidates}, we find three CL-AGN transition candidates among 260 Seyfert AGN over the 6.3-year ZTF baseline. Of these, two are turn-off AGN candidates, leaving us with $f_\text{TO} = 7.8\times10^{-3}$ over 6.3 years. This leads to a calculated AGN episode lifetime of 820 years, roughly four times shorter than the results of \citet{shen2021} for CLQs ($2500 \pm 470$ years). This timescale is consistent with the viscous timescale for black holes at the masses discussed in this work \citep[][]{ricci2022}. However, the actual duration of the changing-look transition ($\sim$years) is far too short for viscous accretion. Further work (e.g. detailed modeling of accretion disk instabilities arising from a lack of fuel to the central engine) will therefore be needed to understand CL-AGN behavior. 

\section{Conclusions}

In this work, we use ZTF light curve data to identify photometric transitions in CL-AGN and Seyfert AGN. We simulate the photometric CL-AGN transition by creating mock photometry with \texttt{sedpy} for the on- and off-state spectra for 23 CL-AGN from the literature. We adopt a threshold of $|\Delta g| > 0.4$ and $|\Delta (g-r)| > 0.2$, which selects 21 of the 23 CL-AGN in this test, as well as one of the two CLQs from \citet{duffy2025} with an observed rise in ZTF. This gives us a completeness of \completeness. The recovered photometric transition rate for 5000 simulated damped random walk light curves is $1.3^{+0.19}_{-0.17}\%$, which we take as our rate of normal AGN activity or flares falling above this threshold.

Adopting this threshold, we find:

\begin{enumerate}
    \item \clagnrate of known CL-AGN experience a photometric transition over the six year ZTF baseline. This is consistent with the predicted rate of photometric transitions given the ZTF coverage of the time between subsequent CL-AGN spectra ($\sim15\%$) and the completeness of our transition criterion.
    \item \syonerate of Seyfert 1s and \sytworate of Seyfert 2s experience a photometric transition over the six year ZTF baseline. The photometric transition rate for Seyfert 1s is consistent with the results of our damped random walk light curve fitting, and therefore we do not expect that these Seyfert 1s are necessarily true CL-AGN. The obscured nature of Seyfert 2s likely accounts for the lack of observed transitions among Seyfert 2s in our sample.
    \item CL-AGN photometric transitions in our sample take a median of 360 days in the rest frame (480 days observer frame), with the shortest transition being 21 days (27 days observer frame) and the longest being 560 days (760 days observer frame). We do not find a correlation between transition amplitude and duration in CL-AGN, possibly due to the small number of photometric transitions identified in this work.
    \item CL-AGN photometric transitions are the most consistent with the thermal and orbital timescales for AGN and may match some theoretical expectations for the timescale of a TDE around an AGN. We do not find a correlation between black hole mass and transition duration, possibly due to the small number of photometric transitions identified in this work.
\end{enumerate}

In addition to these findings, we identify a candidate repeating changing-look event in J0849+2747, a CL-AGN originally identified in \citet{yang2018}. This object has experienced a brightening of 0.59 mag in the $g$-band over 580 days in the rest-frame. Its variability and emission line changes in SDSS are consistent with it being in the process of a repeat turn-on transition $\sim14$ years after it initially turned off. We will investigate this object further in a future work.

The small sample of CL-AGN transition candidates identified in this study limits our ability to identify correlations between CL-AGN transition properties and physical properties of the AGN, which will be necessary in the future to test theories for CL-AGN origins. We have shown that it is possible to monitor large AGN samples in current and future photometric surveys to identify new CL-AGN candidates and photometrically monitor their transitions. Soon, LSST will provide an opportunity to identify CL-AGN photometric transitions across the Southern sky, observing 10,000--1 million CL-AGN through its 10 year survey \citep[][]{lsstsciencebook}. Our pilot study can be expanded using multivariate classifiers \citep[e.g.][]{Breiman2001,chen2016} to identify CL-AGN using LSST light curve data, which will be crucial to confirm or rule out CL-AGN theories discussed in this work.

\begin{acknowledgments}

The authors thank the referees for their feedback which improved this work.

M.E.V. acknowledges support from the Center for Astrophysical Surveys Graduate Fellowship. J.T.H acknowledges support provided by NASA through the NASA Hubble Fellowship grant HST-HF2-51577.001-A awarded by the Space Telescope
Science Institute, which is operated by the Association of Universities for Research in Astronomy, Incorporated, under NASA contract NAS5-26555.

M.E.V. would like to thank Aidan Berres, Gautham Narayan, Haille Perkins, Jessie Runnoe, Yue Shen, and Amanda Wasserman for helpful conversations which improved this work.

Based on observations obtained with the Samuel Oschin Telescope 48-inch and the 60-inch Telescope at the Palomar
Observatory as part of the Zwicky Transient Facility project. ZTF is supported by the National Science Foundation under Grants
No. AST-1440341 and AST-2034437 and a collaboration including current partners Caltech, IPAC, the Oskar Klein Center at
Stockholm University, the University of Maryland, University of California, Berkeley , the University of Wisconsin at Milwaukee,
University of Warwick, Ruhr University, Cornell University, Northwestern University and Drexel University. Operations are
conducted by COO, IPAC, and UW. This work uses data from the ZTF Object Table \citep{ztf_object_table} and ZTF Lightcurves \citep[][]{ztf_light_curves}.

This research has made use of the NASA/IPAC Infrared Science Archive, which is funded by the National Aeronautics and Space Administration and operated by the California Institute of Technology.

Funding for the SDSS and SDSS-II was provided by the Alfred P. Sloan Foundation, the Participating Institutions, the National Science Foundation, the U.S. Department of Energy, the National Aeronautics and Space Administration, the Japanese Monbukagakusho, the Max Planck Society, and the Higher Education Funding Council for England. The SDSS was managed by the Astrophysical Research Consortium for the Participating Institutions.

Funding for the Sloan Digital Sky Survey V has been provided by the Alfred P. Sloan Foundation, the Heising-Simons Foundation, the National Science Foundation, and the Participating Institutions. SDSS acknowledges support and resources from the Center for High-Performance Computing at the University of Utah. SDSS telescopes are located at Apache Point Observatory, funded by the Astrophysical Research Consortium and operated by New Mexico State University, and at Las Campanas Observatory, operated by the Carnegie Institution for Science. The SDSS web site is \url{www.sdss.org}.

SDSS is managed by the Astrophysical Research Consortium for the Participating Institutions of the SDSS Collaboration, including the Carnegie Institution for Science, Chilean National Time Allocation Committee (CNTAC) ratified researchers, Caltech, the Gotham Participation Group, Harvard University, Heidelberg University, The Flatiron Institute, The Johns Hopkins University, L'Ecole polytechnique f\'{e}d\'{e}rale de Lausanne (EPFL), Leibniz-Institut f\"{u}r Astrophysik Potsdam (AIP), Max-Planck-Institut f\"{u}r Astronomie (MPIA Heidelberg), Max-Planck-Institut f\"{u}r Extraterrestrische Physik (MPE), Nanjing University, National Astronomical Observatories of China (NAOC), New Mexico State University, The Ohio State University, Pennsylvania State University, Smithsonian Astrophysical Observatory, Space Telescope Science Institute (STScI), the Stellar Astrophysics Participation Group, Universidad Nacional Aut\'{o}noma de M\'{e}xico, University of Arizona, University of Colorado Boulder, University of Illinois at Urbana-Champaign, University of Toronto, University of Utah, University of Virginia, Yale University, and Yunnan University.  

\end{acknowledgments}

\software{AstroML \citep[][]{astroml,astroMLText}; Astropy \citep{astropy:2013, astropy:2018, astropy:2022}; Matplotlib \citep{matplotlib}; NumPy \citep{numpy}; pandas \citep{pandas,pandas2}; PWLF \citep[][]{pwlf}; PyQSOFit \citep[][]{guo2018,shen2019}; SciPy \citep{scipy}; Seaborn \citep{seaborn}; Sedpy \citep{sedpy}}

\newpage 

\bibliography{clagn}{}
\bibliographystyle{aasjournalv7}

\newpage

\appendix

\section{PyQSOfit Fit Parameters} \label{sec:pyqsofit}

Here we describe the fitting setup we use to derive black hole masses with \texttt{PyQSOfit} \citep[][]{guo2018, shen2019}.

\texttt{PyQSOfit} fits potentially-blended line complexes together when appropriate. We fit all lines which fall in the observed wavelength range for the SDSS spectra for our CL-AGN and Seyfert 1s. These lines include the H$\alpha$ complex (including broad + narrow H$\alpha$, [NII]$\lambda6549$ and $\lambda6585$, and [SII]$\lambda6718$ and $\lambda6732$), H$\beta$ complex, (including broad and narrow H$\beta$, broad and narrow [OIII]$\lambda4959$ and $\lambda5007$, [NII]$\lambda6585$ and a broad and narrow component for HeII$\lambda4687$), CaII$\lambda3934$, [OII]$\lambda3728$, and a broad and narrow component for NeV$\lambda3426$. We also include a polynomial component for dust correction, another for the UV/optical Fe complex, and a host component. 

We fit all lines with a single Gaussian. We use the default priors for line fitting. For broad lines, we allow sigma to vary between $\ln (\sigma_\lambda/\AA)=$ $5\times10^{-3}-0.05$. For narrow lines, we allow sigma to vary between $1\times10^{-3}$ and $1.69\times10^{-3}$ Angstroms in natural log lambda units. We limit velocity offset to 0.01 Angstroms for broad lines and $5\times10^{-3}$ Angstroms for narrow lines in natural log lambda units. We visually inspect all fits for both a good fit to continuum and to the H$\beta$ line. 

We use the MC fitting mode to obtain uncertainties, but propagated uncertainties in black hole mass measurements from line fitting are not significant relative to the intrinsic scatter for the H$\beta-\text{M}_\text{BH}$ relation, so we do not include them. 

\section{Light curves for CL-AGN transition candidates} \label{sec:light_curves}

Here, we present the ZTF light curves for the five known CL-AGN and three Seyfert 1 AGN with a candidate changing-look transition. We overplot the three-segment piecewise linear fit computed with \texttt{\texttt{pwlf}} \citep[][fit described in Section \ref{sec:light_curve_fitting}]{pwlf}. Transition properties are listed in Table \ref{tab:candidate_properties}, where all properties are measured from the middle segment of the fit.

\begin{figure*}
    \centering
    \includegraphics[width=\linewidth]{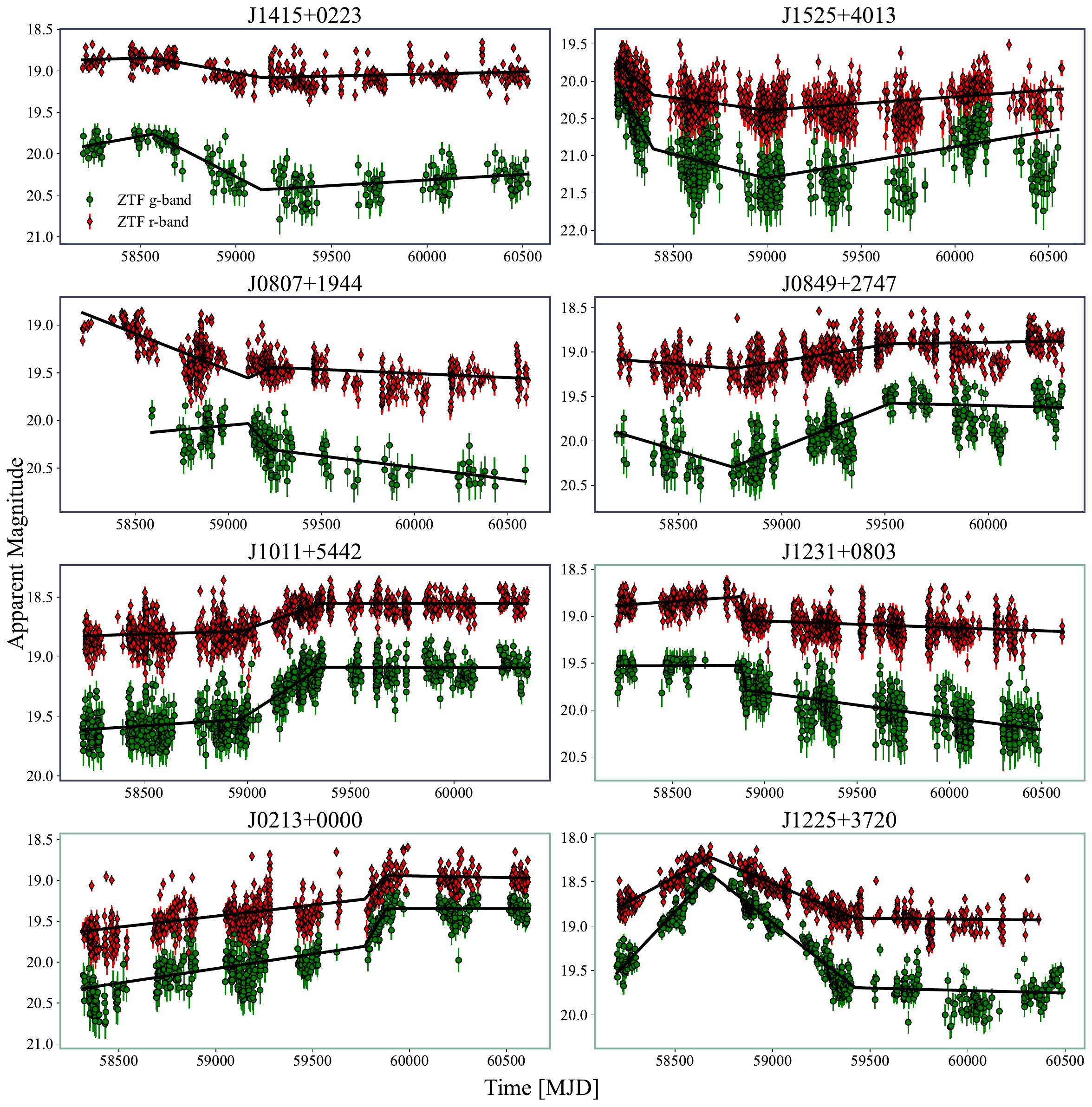}
    \caption{ZTF light curves for the five known CL-AGN (gray borders) and three Seyfert 1 AGN (green borders) with a candidate changing-look transition. Overplotted is the three-segment piecewise linear fit computed with \texttt{pwlf} \citep[][]{pwlf} (black line). Transition properties are listed in Table \ref{tab:candidate_properties}. All CL-AGN except J0849+2747 have behavior consistent with their documented changing-look transition; we identify J0849+2747 as a candidate repeating CL-AGN (see Section \ref{sec:J0849}).}
\end{figure*}

\end{document}